\documentclass[12 pt]{article}
\usepackage{a41}
\usepackage{cite}
\usepackage{axodraw}
\usepackage{color}
\usepackage{epsfig}
\usepackage{amssymb}
\usepackage{amsmath}

\newcommand{\GeV}{ {{\rm GeV}} }
\newcommand{\KI}{ {\mathcal{K}} }

\newcommand{\twz}{ {\text{tw2}} }

\newcommand{\pd}{ \partial }

\newcommand {\PP}{{\mathbb{P}}}

\newcommand {\kln}[1]{\left( #1 \right)}

\newcommand {\KLn}[1]{\Bigl( #1 \Bigr)}

\newcommand {\kls}[1]{\left\{ #1 \right\}}

\newcommand {\kle}[1]{\left[ #1 \right]}

\newcommand {\Matel}[3]{\bigl< #1 \bigl|\; #2\, \bigr| #3 \bigr>}

\newcommand {\im}{{\text{i}}}
\newcommand {\e}{{\text{e}}}

\newcommand\pvec{\mbox{\boldmath $p$}}
\newcommand\lvec{\mbox{\boldmath $l$}}

\begin{document}
\noindent
\sloppy
\thispagestyle{empty}
\begin{flushleft}
{DESY 05--008} \hfill
{\tt hep-ph/0605310}\\
SFB-CPP-06/25\\
May 2006
\end{flushleft}
%
\vspace*{\fill}
\begin{center}

{\LARGE\bf  Target mass and finite momentum transfer }

\vspace{3mm}
{\LARGE\bf  corrections to unpolarized and polarized}

\vspace{3mm}
{\LARGE \bf diffractive scattering}

\vspace{2cm} 
\large Johannes Bl\"umlein$^a$, Bodo Geyer$^b$ and Dieter Robaschik$^{a,c}$ \\ 
\vspace{2em} 
\normalsize {\it $^a$~Deutsches Elektronen--Synchrotron, DESY,\\ Platanenallee 
6, D--15738 Zeuthen, Germany} \\ 

\vspace{2em} $^b${\it Center for Theoretical Studies and 
Institute of Theoretical Physics,\\ Leipzig University, Augustusplatz~10, D-04109~Leipzig, 
Germany} \\ 

\vspace{2em} {\it $^c$~Brandenburgische Technische Universit\"at Cottbus, 
Fakult\"at 1,}\\ {\it PF 101344, D--03013 Cottbus, Germany} \\ 
\end{center}

\vspace*{\fill} 
\begin{abstract} 
\noindent 
A quantum field 
theoretic treatment of inclusive deep--inelastic diffractive scattering is 
given. The process can be described in the general framework of non--forward scattering 
processes using the light--cone expansion in the generalized Bjorken region. Target mass and 
finite $t$ corrections of the diffractive hadronic tensor are derived at the level of the 
twist--2 contributions both for the unpolarized and the polarized case. They modify the 
expressions contributing in the limit $t, M^2 \rightarrow 0$ for larger values of $\beta$ or/and 
$t$ in the region of low $Q^2$. The different diffractive structure functions are expressed
through integrals over the relative momentum of non--perturbative $t$--dependent 2--particle 
distribution functions. In the limit $t, M^2 \rightarrow 0$ these distribution functions are 
the diffractive parton distribution. Relations between the different diffractive structure 
functions are derived.  

\vspace{5mm}
 \end{abstract} 
\vspace*{\fill} 
\newpage 
\section{Introduction} \renewcommand{\theequation}{\thesection.\arabic{equation}} 
\setcounter{equation}{0} 

\vspace{2mm}
\noindent
Deep inelastic diffractive lepton--nucleon scattering was observed at the
electron--proton collider HERA some years ago~\cite{Derrick:1993xh}. This process
is measured in detail by now~\cite{Derrick:1996ma} and the structure function 
$F_2^D(x,Q^2)$ was extracted.~\footnote{The measurement of the longitudinal
diffractive structure function $F_L^D(x,Q^2)$ has not yet been possible, but 
would be important. For the DIS structure function cf.~\cite{Adloff:1996yz}. Likewise the 
polarized diffractive structure functions $g_{1,2}^D(x,Q^2)$ should be measured 
in the future to reveal the effects of nucleon polarization in this process.}
The experimental measurements clearly showed that the scaling violations of the 
deep-inelastic and the diffractive structure functions in the deep-inelastic regime, 
after an appropriate change of kinematic variables, are the same. Furthermore the 
ratio of the two quantities, did not vary strongly, cf. 
\cite{Abramowicz:1996ha}. While the former property is clearly of perturbative nature, 
the latter is of non--perturbative origin.
  
The process of deep--inelastic diffractive scattering was first described phenomenologically
\cite{PHEN}. Diffractive events are characterized by a 
rapidity gap between the diffractive nucleon and the remaining part of the
produced hadrons, which is sufficiently large. Actually it is this experimental 
signature along with factorization for the twist--2 contributions \cite{Berera:1995fj} 
for this process, which allows to give a consistent field theoretic description. 
Due to this phenomenological considerations containing reference to 
specific pomeron models can be thoroughly avoided. 
In the limit of vanishing target masses the 
scattering cross sections and relations between the diffractive structure functions
were derived in Refs.~\cite{Blumlein:2001xf,Blumlein:2002fw,Blumlein:2002ax} 
for unpolarized and polarized diffractive scattering. In \cite{Blumlein:2001xf} we 
showed, that the scaling violations in the deep-inelastic and deeply 
inelastic diffractive case have to be the same due to the fact that the scaling 
violations are actually those of the operators which remain taking the 
respective matrix elements confirming the experimental observation. The 
set of structure functions which 
emerge in both scattering cross sections is actually larger than measured in current 
experiments. At low scales of $Q^2$ target mass effects become relevant similar to the case of
deep--inelastic scattering 
\cite{Nachtmann:1973mr,Georgi:1976ve,Blumlein:1998nv,Blumlein:1999rv,Piccione:1997zh},
see also \cite{TM2}.

In the present paper we extend the picture 
developed in the massless case \cite{Blumlein:2001xf,Blumlein:2002fw} to the case of finite 
target masses and finite values of $t$ on the level of twist--2 operators in the light cone 
expansion \cite{LCE}. While in absence of mass effects the two--particle problem 
could effectively be reduced to a single particle description for the 
case $t \rightarrow 0$, this is no longer the case for finite values
of $t$ and/or target masses. Here two--particle effects become relevant, which do not 
allow for a {\sf direct} partonic description. The variables $t = (p_i - p_f)^2$ and 
$M^2 = p_i^2 = p_f^2$ are closely connected and the simplification emerges if these
scales vanish.
Yet one may still follow the field theoretic picture developed in 
\cite{Blumlein:2001xf,Blumlein:2002fw} in the general case $M^2, t \neq 0$  
and derive expressions for the diffractive structure functions including relations
between them. At low scales $Q^2$ and large values of $\beta$ target mass corrections 
have to be considered in the experimental analysis. This generally applies also to finite 
values of $t$, unless the scale $Q^2$ is large enough. 
The diffractive structure functions are found as  integrals
over two--particle correlation functions $f(z_+,z_-;t)$ between the incoming and 
outgoing nucleon. Here, $z_\pm$ denote the corresponding collinear light--cone 
momentum fractions and $t$ is the relative momentum transfer squared between the 
incoming and outgoing proton momentum. 
We refer to the formalism of non--forward Compton scattering, cf. \cite{Blumlein:1999sc}, 
and apply the general group theoretical algorithm of decomposing off--cone tensor operators 
into operators of definite geometric twist 
\cite{Geyer:1999uq,Eilers:2004mp,E04} to determine the 
contributions at twist 2. The analysis can be generalized to operators of higher twist.
On the level of the various  correlation functions relations can be established. These relations 
correspond to relations between structure functions, 
cf.~\cite{Blumlein:1998nv,Callan:1969uq,Wandzura:1977qf,Blumlein:1996vs,Blumlein:1996tp}. 
In the case of deeply virtual Compton scattering relations of 
this type were found in \cite{Blumlein:2000cx,Geyer:2004bx,Geyer:2004by} before.

The present field--theoretic formalism for deep--inelastic diffractive scattering was
also developed in view of possible future measurements  of the respective 
operator-matrixelements using lattice techniques, as successfully applied in case of 
the moments of the deep--inelastic structure functions~\cite{LAT}. We regard it as a 
challenge in future investigations to verify the experimentally observed ratio of 
moments in the diffractive and deep-inelastic case with these technologies. This 
ratio still awaits a rigorous non--perturbative explanation.

The paper is organized as follows. In section~2 we describe the basic formalism.
In section~3 the symmetric part of the Compton amplitude is dealt with, through which the
diffractive structure functions for unpolarized nucleons are derived. The polarized structure 
functions are determined in section~4. In section~5 we derive relations between different structure 
functions and section~6 contains the conclusions. 
\section{Basic Formalism}
\renewcommand{\theequation}{\thesection.\arabic{equation}}
\setcounter{equation}{0}

\vspace{2mm}
\noindent
In order to compute the twist--2 target mass and finite momentum transfer corrections in 
polarized and
unpolarized deep inelastic diffractive scattering we briefly recall
the notations and conventions used in previous papers \cite{Blumlein:2001xf,Blumlein:2002fw} 
by two of the present authors. The process of deep--inelastic diffractive scattering belongs to 
the class of semi--inclusive processes and is described by the diagram in Figure~1.
The differential scattering cross section for single--photon exchange is given by
\begin{equation}
\label{eqD1}
 \text{d}^5 \! \sigma_{\rm diffr} = \frac{1}{2(s-M^2)} \, \frac{1}{4} \;
           dPS^{(3)} \sum_{\rm spins}\frac{e^4}{Q^2} \, L_{\mu\nu} W^{\mu\nu}~.
\end{equation}
Here $s=\kln{p_1 + l}^2$ is the cms energy squared of the process  and $M$ denotes
the nucleon mass. The phase space $d P S^{\kln{3}}$ depends on five variables
since one final state mass varies. We choose as  basic variables
\begin{equation}
 x = \frac{Q^2}{Q^2 + W^2 - M^2} = - \frac{q^2}{2 \, q p_1} \; ,
\end{equation}
the photon virtuality $Q^2 = - q^2$, 
$t=\kln{p_2 - p_1}^2$ the 4--momentum difference squared between incoming and outgoing 
nucleon, a variable describing the non-forwardness w.r.t.~the incoming
proton direction,
\begin{equation}
\label{eqV1}
x_{\PP} = \frac{Q^2 + M_X^2 - t}{Q^2 + W^2 - M^2} = - \frac{q p_-}{q p_1}
\geq x \;  
\end{equation}
for $M_X^2 > t$, 
and the angle $\Phi $ between the lepton plane $\pvec_1 \times \lvec $ and
the hadron plane  $\pvec_1 \times \pvec_2$,
\begin{equation}
\label{eqV3}
\cos \Phi = \frac{(\pvec_1 \times \lvec).(\pvec_1 \times  \pvec_2)}
                 {|\pvec_1 \times \lvec ||\pvec_1 \times  \pvec_2|}~,
\end{equation}
where $p_\pm = p_2 \pm p_1$, $W^2 = \kln{p_1 + q}^2$ denotes 
the hadronic mass squared and the diffractive mass squared is given by 
$M_X^2 = \kln{q - p_-}^2$.
\\
The momenta $p_\pm$ obey
\begin{equation}
(p_+~p_-) = 0, \qquad \frac{p_+^2}{p_-^2} = \frac{4 M^2}{t} - 1~. 
\end{equation}
For later use we refer to
the variables $\eta$ and $\beta$ defined by
\begin{equation}
 \eta = \frac{q p_-}{q p_+} = \frac{-x_{\PP}}{2-x_{\PP}} 
 \in \kle{-1 \, , \, \frac{-x}{2-x} } \; ,
\qquad
  \beta  =   \frac{q^2}{2\,qp_-} =\frac{x}{x_{\PP}} \leq 1,
\end{equation}
as well as to the transverse momentum variable
\begin{equation}
\pi_- = p_+ - \frac{p_-}{\eta}, \qquad (q \pi_-) = 0~.
\end{equation}
The variables $x, x_\PP, \beta$ and $\eta$ obey the inequalities
\begin{eqnarray}
0 \leq x \leq x_\PP \leq 1, \qquad 0 \leq x \leq \beta \leq 1,\\
-\infty \leq 1 - \frac{2}{x} \leq 1 - \frac{2 \beta}{x} = \frac{1}{\eta} \leq -1 \leq \eta 
\leq \frac{x}{2-x} \leq 0~.
\end{eqnarray}

\vspace*{0.8cm}
\begin{center}
\begin{picture}(-50,100)(0,0)
\setlength{\unitlength}{0.2mm}
\SetWidth{1.5}
\ArrowLine(-150,100)(-100,100)
\ArrowLine(-100,100)(-50,120)
\Photon(-100,100)(-70,70){5}{5}%
\ArrowLine(-30,30)(0,0)
\ArrowLine(-100,0)(-70,30)
\SetWidth{5}
\ArrowLine(-30,70)(0,100)
\SetWidth{1.5}
\CCirc(-50,50){26}{Black}{Yellow}
\setlength{\unitlength}{1pt}
\Text(-160,110)[]{$l$}
\Text(-40,130)[]{$l'$}
\Text(-90,75)[]{$q$}
\Text(-110,-10)[]{$p_1$}
\Text(10,-10)[]{$p_2$}
\Text(10,110)[]{$M_X$}
\end{picture}
\end{center}

\vspace*{1mm}
\noindent
\begin{center}
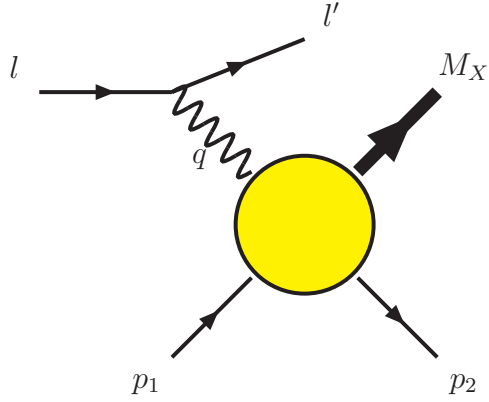
{\sf Figure~1:~The virtual photon-hadron amplitude for diffractive $ep$ scattering} 
\end{center}

\vspace{1mm}\noindent

For the spin averaged cross section, the leptonic tensor is symmetric.
Taking into account conservation of the electromagnetic current one obtains
\cite{Blumlein:2001xf}
\begin{eqnarray}
\label{eqD2}
 W_{\mu\nu}^s
&=&
  -g_{\mu\nu}^T  W_1^s +  p_{1\mu}^T p_{1\nu}^T \frac{W_2^s}{M^2}
   + p_{2\mu}^T  p_{2\nu}^T \frac{W_3^s}{M^2}
   + \kle{ p_{1\mu}^T  p_{2\nu}^T +  p_{2\mu}^T  p_{1\nu}^T } \frac{W_4^s}{M^2} \; .
\end{eqnarray}
Here and in the following we do not assume implicitly, that azimuthal integrals
are performed as sometimes done in experiment. In the latter case the number of 
contributing structure function reduces.

In the case of polarized nucleons we consider the initial state spin--vector 
$S_1\equiv S,~S^2 = -M^2$,  only and sum over the spin of the outgoing 
hadrons. One usually considers the longitudinal $(||)$ and transverse 
($\perp$) spin projections choosing
\begin{eqnarray}
S_{||}    &=& (0;0,0,0,M) \\
S_{\perp} &=& (0;\cos\gamma,\sin\gamma,0) M~.
\end{eqnarray}
Here $\gamma$ denotes the azimuthal angle.
The antisymmetric part of the hadronic tensor was derived in 
\cite{Blumlein:2002fw} and is given by
\begin{alignat}{10}
\label{eqH2}
W_{\mu\nu}^a &=&~~
i \left[ {p}_{1\mu}^T {p}_{2\nu}^T-
{p}_{1\nu}^T {p}_{2\mu}^T \right] \varepsilon_{p_1 p_2 q S}
&\frac{W_1^a}{M^6}&~~
                &+&~~
i \left[ {p}_{1\mu}^T \varepsilon_{\nu S p_1 q}
     - {p}_{1\nu}^T \varepsilon_{\mu S p_1 q} \right]
&\frac{W_2^a}{M^4}&
\nonumber \\ &+&~~
i \left[ {p}_{2\mu}^T \varepsilon_{\nu S p_1 q}
     - {p}_{2\nu}^T \varepsilon_{\mu S p_1 q} \right]
&\frac{W_3^a}{M^4}&~~
                &+&~~
i \left[ {p}_{1\mu}^T \varepsilon_{\nu S p_2 q}
     - {p}_{1\nu}^T \varepsilon_{\mu S p_2 q} \right]
&\frac{W_4^a}{M^4}&
\nonumber \\ &+&~~
i \left[ {p}_{2\mu}^T \varepsilon_{\nu S p_2 q}
     - {p}_{2\nu}^T \varepsilon_{\mu S p_2 q} \right]
&\frac{W_5^a}{M^4}&~~
                &+&~~
i \left[ {p}_{1\mu}^T {\varepsilon}_{\nu p_1 p_2 S}^T
     - {p}_{1\nu}^T {\varepsilon}_{\mu p_1 p_2 S}^T \right]
&\frac{W_6^a}{M^4}&
\nonumber\\ &+&~~
i \left[ {p}_{2\mu}^T {\varepsilon}_{\nu p_1 p_2 S}^T
     - {p}_{2\nu}^T {\varepsilon}_{\mu p_1 p_2 S}^T \right]
&\frac{W_7^a}{M^4}&~~
                &+&~~   i\; \varepsilon_{\mu \nu q S}
&\frac{W_8^a}{M^2}&~.
\end{alignat}
We will specify below which terms of this general structures contribute
in case of deep--inelastic diffractive scattering.
The kinematic factors above are constructed out of the
four--vectors $q,p_1,p_2$ and $S$ as well as $g_{\mu\nu}$
and $\varepsilon_{v_0 v_1 v_2 v_3}$ using
\begin{eqnarray}
p_\mu^T &=& p_\mu - q_\mu \frac{q.p}{q^2},\,\,\,
  g_{\mu\nu}^T=  g_{\mu\nu} - \frac{q_\mu q_\nu}{q^2},          
\\
{\varepsilon}_{\mu v_1 v_2 v_3}^T            &=&
    {\varepsilon}_{\mu v_1 v_2 v_3}            -
    {\varepsilon}_{q v_1 v_2 v_3} \frac{q_\mu}{q^2}~,  \\
{\varepsilon}_{\mu \nu v_1 v_2}^{TT}            &=&
    {\varepsilon}_{\mu \nu v_1 v_2}            -
    {\varepsilon}_{q \nu v_1 v_2} \frac{q_\mu}{q^2}
  - {\varepsilon}_{\mu q v_1 v_2} \frac{q_\nu}{q^2}~.
\end{eqnarray}

At the level of the twist--2 contributions factorization holds for diffractive 
scattering \cite{Berera:1995fj}.  A.~Mueller's generalized optical theorem 
\cite{Mueller:1970fa} allows 
to  move the final state proton into an initial state anti-proton, where both particle
momenta
are separated by $t$ and form a `quasi two--particle' state. The correctness of this 
procedure within the light--cone expansion relies, first, on the rapidity gap between the 
outgoing proton and the remainder hadronic part with invariant mass $M_X$ and, second, on 
the special property of matrix elements of the light-cone operators which 
contain no absorptive part. The structure functions for the diffractive process 
can thus be obtained by analyzing the absorptive part of the expectation value
\begin{equation}
\label{matr}
 T_{\mu\nu}\kln{x} = \Matel{p_1,-p_2, S; t}{ \widehat T_{\mu\nu}(x) }{p_1,-p_2,S; t}~,
\end{equation}
with $\widehat T_{\mu\nu}$ defined as 
\begin{equation}
\label{int_input}
 \widehat T_{\mu\nu}(x) \equiv i R \, T\kle{J_\mu\kln{\frac{x}{2}} \,
                                    J_\nu\kln{-\frac{x}{2}} {\cal S}} \, .
\end{equation}

As shown in \cite{Blumlein:1999sc,Blumlein:2001sb} the operator $\widehat T_{\mu\nu}$ 
is represented 
in lowest order of the non--local light--cone expansion by
\begin{eqnarray}
\label{str_ope}
 \widehat T_{\mu\nu}\kln{x}
&\approx&
 -\, e^2 \frac{\tilde x_\lambda}{ 2\im\pi^2 (x^2-i\epsilon)^2}
 \left[{S_{\mu\nu |}}^{ \alpha\lambda}\,
 O_\alpha   \left(\frac{\tilde x}{2}, -\frac{\tilde x}{2}\right)
- {\epsilon_{\mu\nu}}^{\alpha\lambda}\,
 O_{5\,\alpha} \left(\frac{\tilde x}{2}, -\frac{\tilde x}{2}\right)
\right]~,
\end{eqnarray}
where
\begin{eqnarray}
S_{\mu\nu | \alpha\lambda } = g_{\mu\alpha}g_{\nu\lambda}
                            + g_{\mu\lambda}g_{\nu\alpha}
                            - g_{\mu\nu}g_{\alpha\lambda}~.
\end{eqnarray}
$\tilde x$ denotes a light--like vector related to $x$,
\begin{eqnarray}
\tilde x = x - n [(nx) - \sqrt{(nx)^2 - n^2 x^2}]~, 
\end{eqnarray}
with $n$ a normalized time--like vector,
$n^2 =1$, and the bi--local light--ray
operators $O_\alpha$ and $O_\alpha^5$ read
\begin{eqnarray}
\label{O}
 O_\alpha\kln{\kappa_1 \tilde x,\kappa_2\tilde x} &=&
    i \bigl(\Omega_\alpha\kln{\kappa_1\tilde x,\kappa_2\tilde x}
                    -\Omega_\alpha\kln{\kappa_2\tilde x,\kappa_1\tilde x}\bigr) \\
\label{O5}
O^5_ \alpha\kln{\kappa_1\tilde x,\kappa_2\tilde x} &=&
     \Omega^5_ \alpha\kln{\kappa_1\tilde x,\kappa_2\tilde x}
                    + \Omega^5_ \alpha\kln{\kappa_2\tilde x,\kappa_1\tilde x}~,
\end{eqnarray}
with
\begin{eqnarray}
  \Omega_\alpha\kln{\kappa_1\tilde x,\kappa_2\tilde x}&=&
 R\,T \kln{: \bar\psi\kln{\kappa_1\tilde x} \,\gamma_\alpha\,U\kln{\kappa_1
                  \tilde x,\kappa_2\tilde x}\,\psi\kln{\kappa_2\tilde x} : {\cal S} } \\
 \Omega^5_ \alpha\kln{\kappa_1\tilde x,\kappa_2\tilde x}&=&
   R \, T \kln{: \bar\psi\kln{\kappa_1\tilde x} \, \gamma^5 \gamma_\alpha \,
         U\kln{\kappa_1 \tilde x,\kappa_2\tilde x} \,
         \psi\kln{\kappa_2\tilde x} : {\cal S} }~,
\end{eqnarray}
where $\kappa_1 = - \kappa_2 = 1/2$,~cf.~\cite{Blumlein:1999sc}.
As is well--known, these operators contain contributions of up to twist four 
\cite{Geyer:1999uq}.
The scattering amplitude is obtained by the Fourier transform of the operator
$\widehat T_{\kls{\mu\nu}}\kln{x}$ and forming the matrix element (\ref{matr}).
Here we want to study its twist--2  contributions including the target mass 
and finite momentum transfer corrections.
This is obtained by {\sf harmonic extension} of the twist--2 light-cone 
operators
      $ \Omega^{\rm tw2}_\alpha\kln{\kappa_1 \tilde x,\kappa_2 \tilde x}$
and
      $\Omega^{\rm tw2}_{5\alpha}\kln{\kappa_1 \tilde x,\kappa_2 \tilde x}$
to twist--2 operators
      $ \Omega_\alpha^{\rm tw 2}\kln{\kappa_1 x,\kappa_2 x}$
and
      $\Omega^{\rm tw 2}_{5\alpha}\kln{\kappa_1 x,\kappa_2 x}$
defined off--cone. 
This procedure has been performed for QCD vector operators using group 
theoretical methods
already in Ref.~\cite{Geyer:2001qf} and used for the target mass and 
$t$--corrections of virtual Compton
scattering in Ref.~\cite{Geyer:2004bx}, cf. also \cite{Belitsky:2001hz}. 
A general procedure for determining the complete
(infinite) twist decomposition of off--cone vector operators was introduced in 
Ref.~\cite{Eilers:2004mp}
and recently extended to arbitrary tensor operators in Ref.~\cite{E04}. It 
should be
emphasized that this method works at operator level before taking matrix 
elements.
As a result the twist--2 Compton operator off the light-cone reads 
\cite{Geyer:2001qf,Geyer:2004bx}
\begin{eqnarray}
\label{FTampl}
    \widehat T^{\rm tw2}_{\mu\nu}\kln{q} &=&  \int \, \text{d}^4 \! x \;\,
             \e^{\im qx}\; \widehat T^{\rm tw2}_{\mu\nu}\kln{x}  \\
         &=&
   -\,e^2 \int \frac{d^4 \! x}{2\im\pi^2} \;
      \frac{\e^{\im qx}\,x_{\lambda} }{(x^2- \im\epsilon)^2}
   \{ {S_{\mu\nu |}}^{ \alpha\lambda }\,
    O_\alpha^\twz(\kappa x ,- \kappa x )
      - {\epsilon_{\mu\nu}}^{\alpha\lambda}\,
    O_{5\,\alpha}^{\twz}(\kappa x ,-\kappa x )
    \} \nonumber,
\end{eqnarray}
with
\begin{eqnarray}
\label{NLO}
 {\Omega}^{\rm tw 2}_{(5)\alpha} \kln{\kappa x,-\kappa x}
 =
 \partial_{\alpha} \int_0^1 d\tau 
                 \int \frac{d^4 u}{(2\pi)^4}\,
     {\Omega}_{(5)\mu}(u) \left\{x^\mu (2 + x\pd)
  - \hbox{\large$\frac{1}{2}$}\im\kappa \tau\, u^\mu x^2\right\}
 (3 + x\pd) \mathcal{H}_2 (u, \kappa \tau x)~, \nonumber\\
\end{eqnarray}
cf.~~Eqs. (\ref{O},\ref{O5}),
and
\begin{eqnarray}
\mathcal{H}_\nu (u, \kappa x) = \sqrt{\pi} \,
\Big(\kappa \sqrt{(ux)^2 - u^2 x^2}\Big)^{1/2-\nu}
J_{\nu-1/2}\Bigl(\frac{\kappa}{2} \sqrt{(ux)^2-u^2 x^2}\Bigr)
           e^{i \kappa (xu)/2 }\,,
\label{H2}
\end{eqnarray}
and $\kappa =1 /2$.
In these relations a (formal) Fourier transform of the
non--local operators is introduced and used off--cone later, i.e.,
\begin{eqnarray}
\label{O^Gint} {\Omega}_{(5)\mu}(\kappa \tilde x, -\kappa \tilde x
          )|_{\tilde x \rightarrow x} =
        \frac{1}{(2 \pi)^4}\int d^4 u \,{\Omega}_{(5)\mu}(u)\, 
        \e^{i\kappa (xu)}~.
\end{eqnarray}
Let us emphasize first that (\ref{NLO}) is an off--cone operator 
equation which
determines the twist--2 part of ${\Omega}_{(5)\mu}( \kappa x, -\kappa  x)$
by acting on $\e^{i\kappa (xu)}$ with a corresponding twist--2 projection 
operator
$P^{(2)\mu}_\alpha(x, \partial_x)$ whose result is presented above. 
Second, it
has been shown in \cite{Geyer:2001qf} that, due to this projection and the 
structure of the
Fourier kernel, from the operator ${\Omega}_{(5)\mu}(u)$ in (\ref{NLO}) 
only its
twist--2 part remains.
   
To calculate the twist--2  part of the Compton amplitude 
$T_{\mu\nu}\kln{q}$ we have
to parameterize the non--perturbative expectation values of the twist--2 
operators
$O^{\twz}_{(5)\mu}( \kappa x, -\kappa  x)$. Here, following the general 
approach of \cite{Geyer:2001qf}, this is done by requiring  
\begin{align}
\label{non1}
&\langle p_1, -p_2| e^2\,\im \left(\Omega_{\mu}(u)- 
\Omega_{\mu}(-u)\right)|p_1, -p_2\rangle \nonumber\\
& \hspace{4cm}
= \sum_{a} \KI^a_{\mu}(p_\pm) \int D{\mathbb Z}\,
    \delta^{(4)}(u - p_-z_- - p_+ z_+)\,
    f_{\,a}(z_- ,z_+,t)\,,\\
\label{non15}
&   \langle p_1, -p_2,S | e^2\left(\Omega_{5\mu}(u) + 
\Omega_{5\mu}(-u)\right)|p_1, -p_2,S \rangle \nonumber\\
& \hspace{4cm}
= \sum_{a} \KI^a_{5\,\mu}(p_\pm,S) \int D{\mathbb Z}\,
    \delta^{(4)}(u - p_-z_- - p_+ z_+)\,
    f_{5\,a}(z_- ,z_+,t)\,,
\end{align}
where, in fact, under the $u$-integration the generalized distribution amplitudes
$f_{(5)\,a}(z_- ,z_+; t)$ are reduced to their twist--2 contributions.
These generalized distribution 
amplitudes depend explicitly  on $t$, and additionally there is a
$t-$ and $M^2-$dependence which finally results through the
distribution $\delta^{(4)}(u - p_-z_- - p_+ z_+)$ from the Fourier 
transform in (\ref{FTampl}), together with (\ref{NLO}, \ref{non1}, \ref{non15}), 
cf. Section 3 and 4, as well as from the kinematic pre--factors 
$\KI^a_{(5)\,\mu}(p_\pm,S)$.\footnote{In the 
following the explicit $t$--dependence of the distribution 
functions is understood also when we drop this variables in the respective expressions
for brevity.}
Here, for the symmetric part of the hadronic tensor
we choose $\KI^{a}_{\mu}(p_\pm) $ 
as kinematic pre--factors 
\begin{eqnarray}
\label{kinsym}
\KI^{1 \, \mu} = p_-^\mu, \qquad \qquad
\KI^{2 \, \mu} = \pi_-^\mu \equiv p^\mu_+ - \frac{p^\mu_-}{\eta}~. 
\qquad
\end{eqnarray}
For the antisymmetric case the kinematic factors
\begin{eqnarray}
\label{kinasym}
\KI^{1\, \mu}_{5} = S^\mu, \qquad
\KI^{2\, \mu}_{5} = p_-^\mu \,\frac{(p_2 S)}{M^2}, \qquad
\KI^{3\, \mu}_{5} = \pi_-^\mu\,\frac{(p_2 S)}{M^2}~
%
\quad \\
\nonumber
\end{eqnarray}
contribute. The normalization by $M^2$ in (\ref{kinasym}) is arbitrary and has to be arranged
with the definition of the corresponding non--perturbative distribution functions.
Furthermore,  $ { f}_{(5)\,a}\kln{z_- ,\, z_+; t} \equiv  
f_{(5)\,a}\kln{\mathbb Z; t}$
 denote the respective 2--particle amplitudes
characterized by
the two fractions $z_{\pm}$ of momenta $p_\pm$ 
and a relative kinematic separation in
$t$. The momentum fractions of the incoming $(z_1)$ and outgoing $(z_2)$ 
nucleon are
formally unified into a 2--vector,
\begin{equation}
\mathbb Z = (z_+, z_-)=((z_2+z_1)/2,(z_2-z_1)/2)
\end{equation}
with the measure
\begin{eqnarray}
 D{\mathbb Z} = 2\, dz_+ dz_-\,
 \theta(1-z_++z_-)\,\theta(1+z_+-z_-)\,
 \theta(1-z_+-z_-)\,\theta(1+z_++z_-)\,.
 \end{eqnarray}
Due to the $\delta$--distribution in  (\ref{non1}, \ref{non15}) and the 
structure of the
Compton amplitude, which is determined by 
(\ref{FTampl}--\ref{H2}),
it appears convenient also to unify the momenta as
\begin{eqnarray}
\mathbb P &\equiv& (p_+, p_-) = (p_2 + p_1, p_2 - p_1)
\end{eqnarray}
and to abbreviate 
expressions according to
\begin{eqnarray}
 \Pi^\mu &=& \kappa\,{\mathbb{P^\mu Z}}
                      = \kappa\,(  p_+^\mu z_+ + p_-^\mu z_-) ,
\quad
 \Pi_\mu^{\mathrm T} = g_{\mu\nu}^{\mathrm T} \Pi^\nu
                          = \Pi_\mu - q_\mu \frac{(q\Pi)}{q^2}~,
\label{abk1}
\end{eqnarray}
which replaces $\kappa u$ after performing the $u$-integration.
Note that, due to the square roots in $\mathcal{H}_2(\kappa u, x)$, the 
Fourier transform
can only be performed if $\Pi^2 > 0 $ since  $\sqrt{(q\Pi)^2 - 
q^2\Pi^2}$ has to be real.
In fact, this is equivalent to the requirement 
\begin{eqnarray}
z_+^2 + \frac{t z_-^2}{(4M^2 - t)} \geq 0~, 
\end{eqnarray}
which restricts the allowed values of $z_\pm$ 
as a support condition to the generalized parton distributions 
$f_{(5)\, a}(\mathbb Z, t)$.
It is convenient to introduce a common scale
$\vartheta$ for the integration variables
$z_+ = \vartheta \zeta, \; z_- = \vartheta (1 - \zeta/\eta)$, i.e., to 
change them non-linearly into $\vartheta$ and $\zeta$,
\begin{eqnarray}
    \vartheta = z_- + \frac{z_+}{\eta},
    \qquad
    \zeta = \frac{z_+}{\vartheta} \,.
\end{eqnarray}
Clearly, the $|t|$ values emerging in practice are not expected to deviate
from this condition due to the presence of a non--perturbative damping factor 
$\exp(-b|t|)$ in the distribution functions, with $b \simeq 4...8~\GeV^{-2}$.
Connected to the support condition the positivity of the variable $\Pi^2$
(\ref{pisq}) is guaranteed which is essential for performing Fourier transforms.
The measure  $D{\mathbb Z}$ in these new variables reads
\begin{eqnarray}
 D{\mathbb Z}
 = 2\,|\vartheta|\,d \vartheta d\zeta\,
\theta\big(1-\vartheta+(1+1/\eta)\,\vartheta \zeta \big)
     \theta\big(1+\vartheta-(1+1/\eta)\,\vartheta \zeta\big)&&
 \nonumber \\
\vspace{-3mm} \theta\big(1-\vartheta-(1-1/\eta)\,\vartheta \zeta \big)
    \theta\big(1+\vartheta+(1-1/\eta)\,\vartheta\zeta\big)&&~.
\label{meas1}
\end{eqnarray} 
The parameterization of the matrix elements (\ref{non1}){ and 
(\ref{non15}) } is given by
\begin{eqnarray}
\label{non2}
 {\langle p_1,-p_2 |e^2\,\im\left(\Omega_{\mu}(u)
 - \Omega_{\mu}(u)\right)|p_1,-p_2\rangle}
 &=& {\sum_{a}  \KI^{a }_{\mu}(p_\pm) \int  D{\mathbb Z}\,
   \delta\kln{u - \vartheta {\cal P}} \;
  f_{a}\kln{\vartheta ,{\zeta}{; t} }}\,,\\
\label{non25}
  {\langle p_1,-p_2, S |e^2\left(\Omega_{5\,\mu}(u)
 + \Omega_{5\,\mu}(u)\right)|p_1,-p_2, S\rangle}
 &=& {\sum_{a}  \KI^{a }_{5\,\mu}(p_\pm,S) \int  D{\mathbb Z}\,
   \delta\kln{u - \vartheta {\cal P}} \;
   f_{5\,a}\kln{\vartheta ,{\zeta}{; t} }}\,.
\nonumber\\
\end{eqnarray}

In the following we make the $\vartheta$--dependence of the kinematics
explicit~:
\begin{eqnarray}
\Pi^\mu &=& \kappa \vartheta \,{\cal P}^\mu(\eta;\zeta), \\
{\cal P}^\mu(\eta,\zeta) &=& p^\mu_- + \pi^\mu_- \zeta  
= p^\mu_- (1 - \zeta/{\eta}) + p^\mu_+ \zeta,\\
q\Pi &=& \kappa \vartheta\, (q{\cal P}),\\ 
(q{\cal P}) &=&  (qp_-) = q^2 / 2\beta \leq 0, \\
\label{pisq}
   \Pi^2 &=&  (\kappa \vartheta)^2 {\cal P}^2 ,\\
   {\cal P}^2 &=&  t(1-\zeta/{\eta})^2  +(4M^2-t)\;\zeta^2 \geq 0,
        \\
   (\Pi^{\mathrm T})^2 &=& (\kappa \vartheta)^2 ({\cal P}^{\mathrm T})^2
=  \left(\kappa \vartheta\right)^2 \frac{\left(q {\cal P}\right)^2
   -  q^2 {\cal P}^2}{-q^2} 
\\
     \frac{{\cal P}^2}{({\cal P}^{\mathrm T})^2} 
     &=&  \frac{4 \beta^2 {\cal P}^2 /Q^2}{1+4 \beta^2 {\cal P}^2 /Q^2}
     \geq 0~.
\end{eqnarray}
$\mathcal{P}^2$ depends on both $t$ and the integration variable $\zeta$ 
which is limited
by the Heaviside functions included in the measure (\ref{meas1}). 

Recall, that ${\cal P}^2 \geq 0$ is required which for $t=0$ is 
trivially fulfilled.
Concerning the region of small values of $t  <  0$ one observes 
that $\zeta =0$ is to be excluded. 

As will be shown in Sections 3 and 4, the absorptive part of the 
Compton amplitude
solely emerges from a factor ${1}/[{R(1)+ \im\epsilon}]$ with the 
polynomial
\begin{eqnarray}
 R(\tau) = (q + \tau\Pi)^2= \left(\tau^2 \, \Pi^2+2\tau {q\Pi}+ 
{q^2}\right)
         = \Pi^2(\tau -\tilde \xi_+)(\tau - \tilde \xi_-)\,.
\end{eqnarray}
The roots are
\begin{eqnarray}
\label{xi}
   \tilde \xi_\pm
    = \frac{- q\Pi \pm \sqrt{(q\Pi)^2 - q^2\Pi^2}}{\Pi^2}
    = -\frac{q^2}{q\Pi \pm \sqrt{(q\Pi)^2 - q^2\Pi^2}}~,
\end{eqnarray}
leading to
\begin{eqnarray}
 \label{IM}
 \frac{1}{R(\tau)+ \im\epsilon}=
       \frac{1}{2\sqrt{(q\Pi)^2 - q^2\Pi^2}}
      \left(
      \frac{1}{\tau - \tilde \xi_+ + \im \epsilon}
     - \frac{1}{\tau -\tilde \xi_- - \im\epsilon}  \right)~.
\end{eqnarray}
The imaginary part of the Compton amplitude results  thus from 
\begin{eqnarray}
 \label{imR}
 {\mathrm{Im}} \frac{1}{R(1)+ \im\epsilon}=
          -\frac{ \pi}{2}\frac{1}{\sqrt{(q\Pi)^2 - q^2\Pi^2}}
           \left[
           \delta(1 - \tilde \xi_+ ) + \delta(1 - \tilde \xi_- )
           \right]\,.
\end{eqnarray}
Only the term containing the variable $ \tilde\xi_+ $ contributes to the
cross section, whereas there is no contribution due to the term $ \tilde\xi_- $. 
This is seen as follows.
For negative values of $\vartheta$, $1/\eta \leq \vartheta \leq 0$, $(q\Pi) 
\geq 0$, holds
and therefore, in terms of variables $(\vartheta, \zeta)$, 
one obtains
\begin{eqnarray}
\tilde \xi_+ &=& - \frac{2\beta}{\kappa\vartheta}
\frac{1}{1 + \sqrt{1 + 4\beta^2 \mathcal{P}^2 / Q^2 }}
\Longrightarrow  - \frac{\beta}{\kappa \vartheta} \geq 0 \\
\tilde \xi_- &=& - \frac{2\beta}{\kappa\vartheta}
\frac{1}{1 - \sqrt{1 + 4\beta^2 \mathcal{P}^2 / Q^2 }}
\Longrightarrow   \frac{1}{\kappa \beta \vartheta} \frac{Q^2}{{\cal P}^2}
\rightarrow -\infty 
\end{eqnarray}
taking into account the support condition for $\mathcal{P}(\zeta)$. 
Here $\Longrightarrow$ denotes the limit $Q^2 \rightarrow \infty$.
Therefore $ \delta (1 - \tilde\xi_-) $ does not contribute.
In the case of positive $\vartheta$, $0 \leq \vartheta \leq - 1/\eta$, one 
finds
\begin{eqnarray}
\tilde \xi_+ &=& - \frac{2\beta}{\kappa\vartheta}
\frac{1}{1 + \sqrt{1 + 4\beta^2 \mathcal{P}^2 / Q^2 }}
\Longrightarrow   \frac{1}{\kappa \beta \vartheta} \frac{Q^2}{{\cal P}^2} 
\rightarrow + \infty\\
\tilde \xi_- &=& - \frac{2\beta}{\kappa\vartheta}
\frac{1}{1 - \sqrt{1 + 4\beta^2 \mathcal{P}^2 / Q^2 }}
\Longrightarrow - \frac{\beta}{\kappa \vartheta} \leq 0~,
\end{eqnarray}
and neither $ \delta (1 - \tilde\xi_+) $ nor $ \delta (1 - \tilde\xi_-) $
contribute.

Having this in mind, let us rewrite
\begin{eqnarray}
\tilde{\xi}_+ = \frac{\xi}{\vartheta},
\qquad
\xi =  -\frac{2\beta}{\kappa}\, 
\frac{1}{1 + \sqrt{1 + 4\beta^2 \mathcal{P}^2/Q^2}}
 = \frac{1}{2\kappa\,\beta}\, \frac{Q^2/\mathcal{P}^2}
{1 - \sqrt{1 + 4\beta^2 \mathcal{P}^2/Q^2}}
 \label{xi1}
\end{eqnarray}
from which it is evident that in case of diffractive scattering
$-2\beta$ takes the role of Bjorken variable $x_{Bj}$ in deep inelastic 
scattering
as $\mathcal{P}^2$ takes the role of $4M^2$. Furthermore, rewriting
\begin{equation}
\delta (1 - \tilde\xi_+) = |\vartheta | \delta (\vartheta - \xi) 
\end{equation}
it is obvious that after integration over $\vartheta$ that variable 
is replaced by $\xi$ everywhere. For intermediate momenta $\xi$ 
substitutes the scaling variable $\beta$ and plays the role of a 
generalized Nachtmann variable. Let us note that this variable, however, 
implicitly depends on $\zeta, \beta$ and $\eta$  through ${\cal 
P}(\zeta)$.
 
Due to Eqs. (\ref{xi}), the following equalities
\begin{eqnarray}
\label{pidel}
\left(\Pi^2 + q\Pi \right) \delta(1 - \tilde \xi_+)
    = \sqrt{(q\Pi)^2 - q^2\Pi^2}\; \delta(1 -\tilde \xi_+)
    = -\left( q^2 + q\Pi\right) \delta(1 -\tilde \xi_+)
\,
\end{eqnarray}
hold and can be used to simplify some expressions resulting for the 
absorptive part
of the Compton amplitude.
\section{The Symmetric Part of the Amplitude}
\renewcommand{\theequation}{\thesection.\arabic{equation}}
\setcounter{equation}{0}
\label{sec-unp}

\vspace{2mm}
\noindent
In the case of unpolarized scattering the symmetric part of the Fourier transform of
the operator
$x_{\alpha} O_{\beta}^{\twz}\kln{\kappa x, - \kappa x} /
   \kln{x^2 - \im \epsilon}^2$
determines the contribution to the hadronic tensor of diffractive scattering. It reads 
(cf.~\cite{Geyer:2004bx}, Eq. (6.1)),
\begin{eqnarray}
 {\widehat T}^{\twz}_{\{\mu\nu\}}(q) 
&= &  {S_{\mu\nu|}}^{\alpha\beta} \int \frac{d^4 \! x}{2\pi^2} \;
  \frac{ \e^{\im qx}}{\kln{x^2 - \im\epsilon}^2} \,\, x_{\alpha} i e^2
\KLn{ \Omega^{\twz}_\beta \kln{\kappa x, -\kappa x} -
\Omega^{\twz}_\beta \kln{-\kappa x, \kappa x} }
 \nonumber\\
&=& 2 i e^2\, \int_0^1 d\tau \; \kln{1-\tau
 + \tau \ln\tau} \int \frac{d^4 \! u}{\kappa^4\tau^4} \;
 \left(\Omega^\rho \kln{\frac{u}{\kappa\tau}} -
       \Omega^\rho \kln{\frac{-u}{\kappa\tau}}\right)
 \, \partial_\rho^u \times   \nonumber\\
         &&  \Bigg[\,
            \frac{2\, (q^2)^2}{[(q + u)^2 + i\epsilon]^3}\;
            \Big( u_\mu^{\mathrm T} u_\nu^{\mathrm T} 
            - g_{\mu\nu}^{\mathrm T}\,\left(u^{\mathrm T}\right)^2  \Big)
        +\,
        \frac{u^2\,q^2}{[(q + u)^2 + i\epsilon]^2}\; g_{\mu\nu}^{\mathrm T}
        \Bigg].
\label{OS1}
\end{eqnarray}
We have to form matrix elements
$ \langle p_1 -p_2|\widehat T^{\twz}_{\{\mu\nu\}}(q)|p_1,-p_2\rangle $ 
for this expression using the non--perturbative matrix elements (\ref{non2}) of the operator 
$i(\Omega(u)-\Omega(-u))$. This 
representation of the operator matrix element, based on
the generalized parton distribution functions $f_{a}({\mathbb Z})$, has to be inserted
into the Fourier transform of the Compton operator. 

After a straightforward, but tedious calculation
one obtains for the part which contains the factor $1/(R(1)+\im \varepsilon)$, 
see \cite{Geyer:2004bx}, Eq. (6.4), 
\begin{eqnarray}
   T^{\twz}_{\{\mu\nu\}}(q) &=& 
   \frac{q^2}{2} \int D {\mathbb Z} \Bigg\{\frac{q{\cal K}_a}{q\Pi} \bigg[
      g_{\mu\nu}^{\mathrm T}\; F^a_1({\mathbb Z})
      - \frac{\Pi_\mu^{\mathrm T} \Pi_\nu^{\mathrm T}}{\left(\Pi^{\mathrm T}\right)^2}
      \; F^a_2({\mathbb Z}) \bigg]
\nonumber \\
& &
  + \Big( {\frac{q{\cal K}_a}{q\Pi}}
                      - {\frac{\Pi{\cal K}_a}{\Pi^2}} \Big)
   \bigg[ g_{\mu\nu}^{\mathrm T}\; F^a_3({\mathbb Z})
- \frac{\Pi_\mu^{\mathrm T} \Pi_\nu^{\mathrm T}}{\left(\Pi^{\mathrm T}\right)^2}
    \;  F^a_4({\mathbb Z}) \bigg]
 \nonumber \\
& &  
- \bigg(\frac{{\cal K}_{a\mu}^{\mathrm T}\Pi_\nu^{\mathrm T}
    + \Pi_\mu^{\mathrm T} {\cal K}_{a\nu}^{\mathrm T}}{\left(\Pi^{\mathrm T}\right)^2}
    - 2 \frac{q{\cal K}_a}{q\Pi} 
    \frac{\Pi_\mu^{\mathrm T} \Pi_\nu^{\mathrm T}}{\left(\Pi^{\mathrm T}\right)^2}\bigg)
     \; F_5^a({\mathbb Z}) \Bigg\}
\frac{1}{R(1)+ \im\epsilon}\,.
\label{Ts_nonf}
\end{eqnarray}
In principle there occur also terms $\propto 1/(R(0)+\im \varepsilon)$
which, due to the overall factor $q^2$, do not contribute to the imaginary part
and are  therefore omitted here.
Obviously, the whole expression obeys transversality proving gauge invariance
of the complete amplitude.
The functions $ F^a({\mathbb Z}; q,\Pi) $ are
\begin{eqnarray}
\label{F1}
F^a_1({\mathbb Z}) 
&=& f^a({\mathbb Z})
          + \frac{\Pi^2 (q\Pi+ \Pi^2)}{(q\Pi)^2-q^2\Pi^2}
            \int_0^1\frac{d\tau}{\tau^2}\;f^a\Big(\frac{{\mathbb Z}}{\tau}\Big)
          + \frac{[\Pi^2]^2}{(q\Pi)^2-q^2\Pi^2}
            \int_0^1\frac{d\tau}{\tau^2} \int_0^1\frac{d\tau_1}{\tau^3_1}
            \;f^a\Big(\frac{{\mathbb Z}}{\tau\tau_1}\Big),
\nonumber\\ \\
\label{F2}
\hspace*{-3mm}
F^a_2({\mathbb Z})  
&=&  f^a({\mathbb Z})
          + \frac{3\Pi^2 (q\Pi+ \Pi^2)}{(q\Pi)^2-q^2\Pi^2}
            \int_0^1\frac{d\tau}{\tau^2}\;
            f^a\Big(\frac{{\mathbb Z}}{\tau}\Big)
          + \frac{3[\Pi^2]^2}{(q\Pi)^2-q^2\Pi^2}
            \int_0^1\frac{d\tau}{\tau^2}
            \int_0^1\frac{d\tau_1}{\tau^3_1}\;
            f^a\Big(\frac{{\mathbb Z}}{\tau\tau_1 }\Big)
  \nonumber\\
  &=& 3 F^a_1 - 2 f^a({\mathbb Z}) ,
\\
\label{F3}
F^a_3({\mathbb Z}) 
&=&  - \int_0^1\frac{d\tau}{\tau^2} 
    \left[F_1^a\Big(\frac{\mathbb Z}{\tau}\Big) 
        + \frac{(q\Pi)^2}{(q\Pi)^2-q^2\Pi^2}\,F_2^a\Big(\frac{\mathbb Z}{\tau}\Big) 
    \right]
    - \frac{\Pi^2 (q+\Pi)^2}{(q\Pi)^2-q^2\Pi^2}
    \int_0^1\frac{d\tau}{\tau^2}\; f^a\Big(\frac{\mathbb Z}{\tau}\Big) 
    \nonumber\\
&&    + \frac{2\,(q\Pi)}{(q\Pi)^2-q^2\Pi^2}
    \int_0^1\frac{d\tau}{\tau^2}
    \left[(q\Pi + \Pi^2)\; f^a\Big(\frac{\mathbb Z}{\tau}\Big) 
    + \Pi^2\int_0^1\frac{d\tau_1}{\tau_1^2}\;f^a\Big(\frac{\mathbb Z}{\tau\tau_1}\Big)
    \right]
    \nonumber\\
&&  + \frac{\Pi^2}{(q\Pi)^2-q^2\Pi^2}
    \left[(q\Pi + q^2)\; f^a({\mathbb Z}) 
    + (q\Pi)\int_0^1\frac{d\tau}{\tau^2}\;f^a\Big(\frac{\mathbb Z}{\tau}\Big)
    \right],
\\
\label{F4}
F^a_4({\mathbb Z})    
&=&   3 F^a_3 + 2  \frac{\Pi^2 }{(\Pi^{\mathrm T})^2} F^a_5   
    -2 \frac{\Pi^2}{(q\Pi)^2-q^2\Pi^2}
    \left[(q\Pi + q^2)\; f^a({\mathbb Z}) 
    + (q\Pi)\int_0^1\frac{d\tau}{\tau^2}\;f^a\Big(\frac{\mathbb Z}{\tau}\Big)
    \right],
 \\
\label{F5}
F^a_5({\mathbb Z}) 
&=&  \int_0^1\frac{d\tau}{\tau^2}\;
            F_2^a\Big(\frac{{\mathbb Z}}{\tau}\Big)\,.
\end{eqnarray}
The imaginary part of the Compton amplitude is given by
\begin{eqnarray}
{\sf Im}\,T^\twz_{\{\mu\nu\}}\kln{q}
 &=&
 \frac{- \pi}{2 \kappa^2} \int d\zeta \int_{{1/\eta}}^{{0}} {d\vartheta}\,
 \frac{q^2}{[(q{\cal P})^2 - q^2 {\cal P}^2]^{1/2}}
\;\delta(\vartheta-\xi)
\nonumber\\ & & \times
 \Bigg\{
 \frac{q{\cal K}_a}{q{\cal P}}\; \bigg[
 g_{\mu\nu}^{\mathrm T} F^a_1(\vartheta,\zeta)
 -
 \frac{{\cal P}_\mu^{\mathrm T} {\cal P}_\nu^{\mathrm T}}
      {({\cal P}^{\mathrm T})^2}\,
 F^a_2(\vartheta,\zeta) \bigg]
 \nonumber \\
& &
 +
 \Big( \frac{q{\cal K}_a}{q{\cal P}}
     - \frac{{\cal P}{\cal K}_a}{{\cal P}^2} \Big)
 \bigg[
 g_{\mu\nu}^{\mathrm T} F^a_3(\vartheta,\zeta)
 - \frac{{\cal P}_\mu^{\mathrm T} {\cal P}_\nu^{\mathrm T}}
      {({\cal P}^{\mathrm T})^2}\,
 F^a_4(\vartheta,\zeta) \bigg]
\nonumber\\ & & -
 \bigg(\frac{{\cal K}_{a\mu}^{\mathrm T} {\cal P}_\nu^{\mathrm T}
 + {\cal P}_\mu^{\mathrm T} {\cal K}_{a\nu}^{\mathrm T}}
      {({\cal P}^{\mathrm T})^2}
 - 2 \frac{q{\cal K}_a}{q{\cal P}}
      \frac{{\cal P}_\mu^{\mathrm T} {\cal P}_\nu^{\mathrm T}}
      {({\cal P}^{\mathrm T})^2} \bigg)
 F^a_5(\vartheta,\zeta)
\Bigg\}~.
\end{eqnarray}
We change now to the variables $(\vartheta, \zeta)$ and express the above integrals
using the following new distribution amplitudes~:
\begin{eqnarray}
\Phi_a^{(0)}(\vartheta,\zeta) &\equiv& f_a (\vartheta,\zeta),
\nonumber\\
\Phi_a^{(1)}(\vartheta,\zeta) 
&\equiv&
      \int_\vartheta^{1/\eta} dy_1 \;f_a (y_1,\zeta)
 = \vartheta \int_0^1 \frac{d\tau}{\tau^2}\;
   f_a \Big(\frac{\vartheta}{\tau},\zeta\Big),
\\
\label{eqphia2}
\Phi_a^{(2)}(\vartheta,\zeta) 
&\equiv& 
\int_\vartheta^{1/\eta} dy_2 
\int_{y_2}^{1/\eta} dy_1 \, f_a (y_1,\zeta)
= \vartheta^2 \int_0^1 \frac{d\tau_1}{\tau_1^3}
  \int_0^1 \frac{d\tau_2}{\tau_2^2}\;
   f_a \Big(\frac{\vartheta}{\tau_1\tau_2},\zeta\Big),
\\
\Phi_a^{(i)}(\vartheta,\zeta) 
&\equiv& 
\int_\vartheta^{1/\eta} dy\; \Phi_a^{(i-1)}(y,\zeta)~.
\end{eqnarray}
In the following we consider the imaginary part of the Compton amplitude
which forms the hadronic tensor for deep--inelastic diffractive scattering.
As already stated above, the imaginary part
follows from ${\sf Im}\; \{1/(R +i\epsilon)\} $.
The $\vartheta$ integral is easily carried out

\begin{eqnarray}
{\sf Im}\,T^\twz_{\{\mu\nu\}}\kln{q}
 &=&
 \frac{- \pi}{2 \kappa^2} \int d\zeta 
 \frac{q^2}{[(q{\cal P})^2 - q^2 {\cal P}^2]^{1/2}}
\;
\nonumber\\ & & \times
 \Bigg\{
 \frac{q{\cal K}_a}{q{\cal P}}\; \bigg[
 g_{\mu\nu}^{\mathrm T} F^a_1(\xi,\zeta)
 -
 \frac{{\cal P}_\mu^{\mathrm T} {\cal P}_\nu^{\mathrm T}}
      {({\cal P}^{\mathrm T})^2}\,
 F^a_2(\xi,\zeta) \bigg]
 \nonumber \\
& &
 +
 \Big( \frac{q{\cal K}_a}{q{\cal P}}
     - \frac{{\cal P}{\cal K}_a}{{\cal P}^2} \Big)
 \bigg[
 g_{\mu\nu}^{\mathrm T} F^a_3(\xi,\zeta)
 - \frac{{\cal P}_\mu^{\mathrm T} {\cal P}_\nu^{\mathrm T}}
      {({\cal P}^{\mathrm T})^2}\,
 F^a_4(\xi,\zeta) \bigg]
\nonumber\\ & & -
 \bigg(\frac{{\cal K}_{a\mu}^{\mathrm T} {\cal P}_\nu^{\mathrm T}
 + {\cal P}_\mu^{\mathrm T} {\cal K}_{a\nu}^{\mathrm T}}
      {({\cal P}^{\mathrm T})^2}
 - 2 \frac{q{\cal K}_a}{q{\cal P}}
      \frac{{\cal P}_\mu^{\mathrm T} {\cal P}_\nu^{\mathrm T}}
      {({\cal P}^{\mathrm T})^2} \bigg)
 F^a_5(\xi,\zeta)
\Bigg\}\!\!~,
\label{Ts_nonfim}
\end{eqnarray}
where the substitution $\Pi_\mu = \kappa \xi {\cal P}_\mu$ 
was used. Here the functions $F^a_i({\mathbb Z}; q,\Pi)\equiv
F^a_i(\xi ,\zeta;q,\kappa \xi {\cal P}),\,i=1,\ldots,5,$ are obtained
from (\ref{F1}) -- (\ref{F5}) using
 (\ref{pidel}) in a non--trivial way~:
\begin{eqnarray}
\label{F1x}
F^a_1(\xi,\zeta)
  & \equiv &
  \Phi_a(\xi,\zeta)
  + \frac{\kappa {\cal P}^2}{[(q{\cal P})^2-q^2{\cal P}^2]^{1/2}}
  \Phi_a^1(\xi, \zeta)
  + \frac{\kappa^2 [{\cal P}^2]^2}{(q{\cal P})^2-q^2{\cal P}^2}
  \Phi_a^2(\xi, \zeta)
\\
\label{F2x}
F^a_2(\xi,\zeta)
  & \equiv &
  \Phi_a(\xi,\zeta)
  + \frac{3\kappa {\cal P}^2}{[(q{\cal P})^2-q^2{\cal P}^2]^{1/2}}
  \Phi_a^1(\xi,\zeta)
  + \frac{3\kappa^2  [{\cal P}^2]^2}{(q{\cal P})^2-q^2{\cal P}^2}
  \Phi_a^2(\xi,\zeta)
\end{eqnarray}
\begin{eqnarray}
\label{F3x}
F_{a\, 3}(\xi,\zeta)
   &\equiv&
  -\int_0^1\frac{d\tau}{\tau^2}
  \bigg[F_{a\, 1}\Big(\frac{\xi}{\tau},\zeta\Big)
  + \frac{(q{\cal P})^2}{(q{\cal P})^2-q^2{\cal P}^2}\,
    F_{a\, 2}\Big(\frac{\xi}{\tau},\zeta\Big)\bigg]
  \nonumber \\
&&  + \frac{2(q{\cal P})}{[(q{\cal P})^2-q^2{\cal P}^2]^{1/2}}
  \int_0^1\frac{d\tau}{\tau^2}
  \bigg[
  \Phi_a^{(0)}\Big(\frac{\xi}{\tau},\zeta\Big)
  +\frac{\kappa \,{\cal P}^2}{[(q{\cal P})^2-q^2{\cal P}^2]^{1/2}} \,
  \Phi_a^{(1)}\Big(\frac{\xi}{\tau},\zeta\Big)
  \bigg]
  \nonumber \\
&&  +\frac{\kappa \,{\cal P}^2}{[(q{\cal P})^2-q^2{\cal P}^2]^{1/2}}
  \bigg[
  -\,\xi \,\Phi_a^{(0)}(\xi,\zeta)
  +  \frac{q{\cal P}}{[(q{\cal P})^2-q^2{\cal P}^2]^{1/2}} \,
  \Phi_a^{(1)}(\xi,\zeta)
  \bigg],
\\
\label{F4x}
F_{a\, 4}(\xi,\zeta)
    &\equiv&
  3 F_{a\, 3}(\xi,\zeta) 
  -  \frac{2\,q^2{\cal P}^2}{[(q{\cal P})^2-q^2{\cal P}^2]}\,
    F_{a\, 5}(\xi,\zeta) 
  \nonumber \\
&& -   \frac{2\,\kappa \,{\cal P}^2}{[(q{\cal P})^2-q^2{\cal P}^2]^{1/2}}
  \bigg[  
  -\,\xi \,\Phi_a^{(0)}(\xi,\zeta)
  +  \frac{q{\cal P}}{[(q{\cal P})^2-q^2{\cal P}^2]^{1/2}} \,
   \Phi_a^{(1)}(\xi,\zeta)
  \bigg] ,
\\
F_{a\, 5}(\xi,\zeta)
  &\equiv&
  \int_0^1\frac{d\tau}{\tau^2}\,
  F_{a\, 2}\Big(\frac{\xi}{\tau},\zeta\Big)\,.
  \label{F5x}
\end{eqnarray}
In the limit $t,~M^2~\rightarrow~0$, Eqs.~(5,8) of Ref.~\cite{Blumlein:2001xf},  
the hadronic tensor
simplifies and contains only two structure functions which are linear combinations of the
four  occurring in the general case (\ref{eqD2}) weighted by powers of
$(1-x_{\PP})$. In this limit the  functions $\left. F_k^a\right|_{k=3}^5$ do not vanish
and contribute to the two structure functions remaining.
On the other hand, $\left. F_k^a\right|_{k=3}^5$ do not contribute for forward scattering
$p_2 \rightarrow 0$.

Now, we can rewrite the imaginary part of the Compton amplitude, simply having 
recourse to  \cite{Geyer:2004bx}, Eqs. (6.27) and (6.36), and replacing 
there $x \rightarrow -2\beta$~:
\begin{eqnarray}
\label{kd1}
{\sf Im}\,T^{\twz}_{\{\mu\nu\}}\kln{q}=
   {\sf Im}\,T_{1\,\{\mu\nu\} }^{\twz}\kln{q} +
    {\sf Im}\,T_{2\,\{\mu\nu\}}^{\twz}\kln{q}~,
\end{eqnarray}
where
\begin{eqnarray}
{\sf Im}\, T^\twz_{1\,\{\mu\nu\}}\kln{q}
 &=&\sum_a 2 \pi \int d\zeta\;
   \frac{q{\cal K}_a}{q\cal P}\,
  \bigg[
 - g_{\mu\nu}^{\mathrm T}\, {W}^{\rm diff}_{a\,1} (\xi, \beta, \eta;\zeta)
 + \frac{{\cal P}_\mu^{\mathrm T} {\cal P}_\nu^{\mathrm T}}{{\cal P}^2}
 \,{W}^{\rm diff}_{a\,2}(\xi, \beta, \eta;\zeta)\bigg]\,
 \label{Ts_end}\\
 {\sf Im}\, T^\twz_{2\,\{\mu\nu\}\,}\kln{q}
 &=&\sum_a 2 \pi  \int d\zeta\;
 \Bigg\{
 \Big( \frac{q{\cal K}_a}{q{\cal P}}
     - \frac{{\cal P}{\cal K}_a}{{\cal P}^2} \Big)
 \bigg[
 g_{\mu\nu}^{\mathrm T}
 - 3\,\frac{{\cal P}_\mu^{\mathrm T}{\cal P}_\nu^{\mathrm T}}{({\cal P}^{\mathrm T})^2}
  \bigg] 
  \nonumber\\ 
& & \times
  \bigg[
  \int_{\xi}^{1/\eta} \frac{d y}{\xi}
  \bigg(
   {W}^{\rm diff }_{a\,1}(y,\beta,\eta;\zeta)
  + \frac{Q^2}{4\beta^2{\cal P}^2}\,
   {W}^{\rm diff }_{a\,2}(y,\beta ,\eta;\zeta)
   \bigg)
  \nonumber \\
& & ~ + 
  \int_{\xi}^{1/\eta} \frac{d y}{\xi}
  \bigg(
  2 V^{\rm diff }_{a\,0}(y,\beta, \eta;\zeta)
 +
  V^{\rm diff }_{a\,1}(y,\beta, \eta;\zeta)
  \bigg) 
  \bigg]
  \nonumber \\
& & +\; 
 \Big({\cal P}_\mu^{\mathrm T}g_{\nu\alpha}^{\mathrm T}
       + {\cal P}_\nu^{\mathrm T}g_{\mu\alpha}^{\mathrm T} 
       - 2\,
 \frac{{\cal P}_\mu^{\mathrm T}{\cal P}_\nu^{\mathrm T}}{({\cal P}^{\mathrm T})^2} 
 {\cal P}^{\mathrm T}_\alpha \Big)
 \frac{{\cal K}^{\alpha}_a}{{\cal P}^2}
  \int_{\xi}^{1/\eta} \frac{d y}{\xi}
  {W}^{\rm diff }_{a\,2}(y,\beta ,\eta;\zeta)
 \nonumber \\
& & 
  +\;\frac{1}{2} \Big( \frac{q{\cal K}_a}{q{\cal P}}
     - \frac{{\cal P}{\cal K}_a}{{\cal P}^2} \Big)
 \bigg[
 g_{\mu\nu}^{\mathrm T}
 - \,\frac{{\cal P}_\mu^{\mathrm T}{\cal P}_\nu^{\mathrm T}}{({\cal P}^{\mathrm T})^2}
  \bigg] V^{\rm diff }_{a\,1}(\xi,\beta,\eta;\zeta) 
\Bigg\}\,,
\label{TSext}
\end{eqnarray}
where
\begin{align}
     {W}^{\rm diff}_{a\,1 }(\xi, \beta,\eta; \zeta)
  & \equiv
    \frac{2\beta}{\sqrt{1+4\beta^2\mathcal{P}^2/Q^2}}
    \; F^a_1(\xi,\beta,\eta; \zeta),
\label{W10}
    \\
     {W}^{\rm diff}_{a\,2 }(\xi, \beta,\eta; \zeta)
  & \equiv
    \frac{(2\beta)^3}{(\sqrt{1+4\beta^2\mathcal{P}^2/Q^2})^3}
    \;F^a_2(\xi,\beta,\eta; \zeta) 
\label{W20}
\end{align}
and
\begin{align}
     V^{\rm diff}_{a\,0}(\xi, \beta,\eta; \zeta)
  & \equiv
    \frac{-2\beta\,\Phi_a^{(0)}(\xi;\zeta)}{\sqrt{1+4\beta^2\mathcal{P}^2/Q^2}}\,,
\label{H0}
    \\ 
\label{va1}
    V^{\rm diff}_{a\,1}(\xi,\beta,\eta;\zeta)
  & \equiv
    \frac{-2\beta^2\mathcal{P}^2}{Q^2}\frac{\partial}{\partial \beta}\left(
    \frac{-2\beta\,\Phi_a^{(1)}(\xi;\zeta)}{\sqrt{1+4\beta^2\mathcal{P}^2/Q^2}}
    \right)
    \\   &  
    =
    \frac{4\beta^2\mathcal{P}^2/Q^2}{1+4\beta^2\mathcal{P}^2/Q^2}
    \left[
     -\xi\, \Phi_a^{(0)}(\xi,\zeta)
    +
    \frac{\Phi_a^{(1)}(\xi,\zeta)}{\sqrt{1+4\beta^2\mathcal{P}^2/Q^2}}
    \right]. \nonumber
\end{align}
Note that $\xi=\xi(\beta,\eta;\zeta)$ and $\mathcal{P}=\mathcal{P}(\eta;\zeta)$.
Differential operators as emerging in (\ref{va1}) are characteristic for 
target mass corrections and were applied earlier in 
Refs.~\cite{Georgi:1976ve,Blumlein:1998nv},
see also \cite{Blumlein:1996vs}. If the corresponding differential expressions are found the associated 
integral representations are derived straightforwardly.

We now convert the result referring to bases like used in (\ref{eqD2}).
By observing 
\begin{eqnarray}
\frac{{\cal P}^{\mathrm T}{\cal K}^{\mathrm T}_a}{{\cal P}^2}
= \frac{{\cal P}{\cal K}_a}{{\cal P}^2}
+ \frac{1}{4\beta^2{\cal P}^2/Q^2}\frac{q{\cal K}_a}{q{\cal P}},
\end{eqnarray}
one obtains
\begin{eqnarray}
\frac{1}{2\pi}\;{\sf Im}\,T_{\{\mu\nu\}}\kln{q}  
&=& 
\sum_a \int d\zeta
\bigg[
    - g_{\mu\nu}^{\mathrm T}
    \left(\frac{q{\cal K}_a}{q\cal P}\; T_1^a(\xi,\zeta)
        + \frac{{\cal P}{\cal K}_a}{{\cal P}^2}\;T_2^a(\xi,\zeta)\right)
\\ & &
    + \frac{{\cal P}_\mu^{\mathrm T}{\cal P}_\nu^{\mathrm T}}{({\cal P}^{\mathrm T})^2}
    \left(\frac{q{\cal K}_a}{q\cal P}\; T_3^a(\xi,\zeta)       
        + \frac{{\cal P}{\cal K}_a}{{\cal P}^2}\; T_4^a(\xi,\zeta)\right)
    + \frac{{\cal K}_{a\,\nu}^{\mathrm T}{\cal P}_\mu^{\mathrm T}
          + {\cal K}_{a\,\mu}^{\mathrm T}{\cal P}_\nu^{\mathrm T}}{{\cal P}^2}\;
     T_5^a(\xi,\zeta) 
\bigg]. \nonumber
\label{kinsymTp}
\end{eqnarray}
The  distribution functions $ T_i^a(\xi,\zeta)$ are related to the distributions used above by
\begin{align}
\label{T1}
T_1^a(\xi,\zeta) 
&=  
{W}^{\rm diff }_{a\,1}(\xi, \beta,\eta;\zeta) - T_2^a(\xi,\zeta) \,, 
\\
T_2^a(\xi,\zeta) 
&=
  \int_{\xi}^{1/\eta} \frac{dy}{\xi}
  \left[2\,{W}^{\rm diff}_{a\,L} - {W}^{\rm diff}_{a\,2}
        +\,V^{\rm diff }_{a\,1}\right]\!
  (y,\beta, \eta;\zeta) 
+\,\frac{1}{2}\, V^{\rm diff }_{a\,1}(\xi, \beta,\eta;\zeta)\,,
 \label{T2}\\
T_3^a(\xi,\zeta) 
&=   
 \big(W^{\rm diff }_{a\,1}+W^{\rm diff }_{a\,L}\big)(\xi,\beta,\eta;\zeta)
- \frac{Q^2}{2\beta^2{\cal P}^2} 
  \int_{\xi}^{1/\eta} \frac{dy}{\xi}
   {W}^{\rm diff }_{a\,2}(y,\beta,\eta ;\zeta) 
   \nonumber\\ & -\Big(3\, T_2^a(\xi,\zeta)
   -\, V^{\rm diff }_{a\,1}(\xi,\beta,\eta;\zeta)\Big)\,,
\label{T3} \\
T_4^a(\xi,\zeta) 
&= - 2
  \int_{\xi}^{1/\eta} \frac{dy}{\xi}
   {W}^{\rm diff }_{a\,2}(y,\beta,\eta ;\zeta)
   +
   \Big(3\,T_2^a(\xi,\zeta)
   -\, V^{\rm diff }_{a\,1}(\xi,\beta,\eta;\zeta)\Big)
    \,,
\label{T4} \\
  \label{T5}
T_5^a(\xi,\zeta) 
&= 
  \int_{\xi}^{1/\eta} \frac{dy}{\xi}
  {W}^{\rm diff }_{a\,2}(y,\beta ,\eta;\zeta) \,.
\end{align}
The $\beta$ and $\eta$ dependence of $T^a_i(\xi,\zeta)$ is understood.
There are different possibilities to choose the tensor--basis of the hadronic tensor.
One way is
\begin{eqnarray}
\label{kinsymTp1} \frac{1}{2\pi}\,{\sf Im}\,T_{\{\mu\nu\}}\kln{q}
       &=& -g_{\mu\nu}^{\mathrm T}
       \int d\zeta\sum_a
       \Big[
         \frac{q{\cal K}_a}{q\cal P}\, T_1^a(\xi,\zeta)
       + \frac{{\cal P}{\cal K}_a}{{\cal P}^2}\, T_2^a(\xi,\zeta) 
       \Big] 
\nonumber \\
       && +\frac{1}{M^2}
       p_{-\mu}^{\mathrm T} p_{-\nu}^{\mathrm T} 
       \int d\zeta \,\frac{M^2}{{({\cal P}^{\mathrm T})^2}} \sum_a
       \Big[
         \frac{q{\cal K}_a}{q\cal P}\, T_3^a(\xi,\zeta)
       + \frac{{\cal P}{\cal K}_a}{{\cal P}^2}\, T_4^a(\xi,\zeta)
       \Big] 
\nonumber 
\end{eqnarray}
\begin{eqnarray}
       && +\frac{1}{M^2}
       (p_{-\mu}^{\mathrm T} \pi_{-\nu}^{\mathrm T}
       +p_{-\nu}^{\mathrm T} \pi_{-\mu}^{\mathrm T})
       \int d\zeta\, \zeta \,\frac{M^2}{{({\cal P}^{\mathrm T})^2}} \sum_a
       \Big[
         \frac{q{\cal K}_a}{q\cal P}\, T_3^a(\xi,\zeta)
       + \frac{{\cal P}{\cal K}_a}{{\cal P}^2}\, T_4^a(\xi,\zeta)
       \Big]             
\nonumber \\
       && + \frac{1}{M^2}
       \pi_{-\mu}^{\mathrm T} \pi_{-\nu}^{\mathrm T} 
       \int d\zeta\, \zeta^2 \,\frac{M^2}{{({\cal P}^{\mathrm T})^2}} 
       \sum_a 
       \Big[
       \frac{q{\cal K}_a}{q\cal P}\, T_3^a(\xi,\zeta)
       + \frac{{\cal P}{\cal K}_a}{{\cal P}^2}\, T_4^a(\xi,\zeta)
       \Big] 
\nonumber \\
       &&  + \frac{1}{M^2}
       (p_{-\mu}^{\mathrm T} {\cal K}_{a\nu}^{\mathrm T}
       +p_{-\nu}^{\mathrm T} {\cal K}_{a\mu}^{\mathrm T})
       \int d\zeta \,\frac{M^2}{{\cal P}^2} \sum_a  T_5^a(\xi,\zeta) 
\nonumber \\
       &&  + \frac{1}{M^2}
       (\pi_{-\mu}^{\mathrm T}{\cal K}_{a\nu}^{\mathrm T}  
       +\pi_{-\nu}^{\mathrm T}{\cal K}_{a\mu}^{\mathrm T})
       \int d\zeta\, \zeta\,\frac{M^2}{{\cal P}^2} \sum_a T_5^a(\xi,\zeta)~.
\end{eqnarray}
Here 
the two kinematic invariants are ${\cal K }^{1\,\rho}= p_-^\rho,\,\,
{\cal K }^{2\,\rho}= \pi_-^\rho$. Introducing the structure functions $U_i(\beta,\eta)$ one  
obtains
\begin{eqnarray}
\label{kinsymTp2}
\frac{1}{2\pi}\, {\sf Im}\,T_{\{\mu\nu\}}\kln{q} & =&
        - g_{\mu\nu}^{\mathrm T}\;U_1(\beta,\eta)
        + \frac{ p_{-\mu}^{\mathrm T} p_{-\nu}^{\mathrm T}}{M^2} \; U_2(\beta,\eta)
         \nonumber\\        &&   
        + \frac{ p_{-\mu}^{\mathrm T} \pi_{-\nu}^{\mathrm T} +
                 p_{-\nu}^{\mathrm T} \pi_{-\mu}^{\mathrm T}}{M^2}\;U_3(\beta,\eta)
        + \frac{\pi_{-\mu}^{\mathrm T}\pi_{-\nu}^{\mathrm T}}{M^2}\;U_4(\beta,\eta)
\end{eqnarray}
with
\begin{eqnarray}
\label{kinsymTp3}
  U_1(\beta,\eta)&=&\int d\zeta
            \Big[T_1^1(\xi,\zeta)
            + \frac{{\cal P}p_-}{{\cal P}^2} T_2^1(\xi,\zeta)
            + \frac{{\cal P}\pi_-}{{\cal P}^2} T_2^2(\xi,\zeta)
            \Big]\,, 
\\
  U_2(\beta,\eta)&=&\int d\zeta \bigg\{
             \frac{M^2}{({\cal P}^{\mathrm T})^2}
             \Big[T_3^1(\xi,\zeta)
             + \frac{{\cal P}{p_-}}{{\cal P}^2}\,T_4^1(\xi,\zeta)
             + \frac{{\cal P}{\pi_-}}{{\cal P}^2}\,T_4^2(\xi,\zeta)\Big]
             + \frac{M^2}{{\cal P}^2}\;T_5^1(\xi,\zeta) 
            \bigg\}, \nonumber\\ 
\end{eqnarray}\begin{eqnarray}
  U_3(\beta,\eta)&=&\int d\zeta\, \zeta  \bigg\{
             \frac{M^2}{({\cal P}^{\mathrm T})^2}
             \Big[T_3^1(\xi,\zeta)
             + \frac{{\cal P}{p_-}}{{\cal P}^2}\,T_4^1(\xi,\zeta)
             + \frac{{\cal P}{\pi_-}}{{\cal P}^2}\,T_4^2(\xi,\zeta)\Big]
             \nonumber\\ & & 
             + \frac{M^2}{{\cal P}^2} \Big[T_5^1(\xi,\zeta)   
             +\,\frac{1}{\zeta}\,T_5^2(\xi,\zeta)
             \Big]\bigg\}\,,
\\
  U_4(\beta,\eta)&=&\int d\zeta\, \zeta^2  \bigg\{
             \frac{M^2}{({\cal P}^{\mathrm T})^2}
             \Big[T_3^1(\xi,\zeta)
             + \frac{{\cal P}{p_-}}{{\cal P}^2}\,T_4^1(\xi,\zeta)
             + \frac{{\cal P}{\pi_-}}{{\cal P}^2}\,T_4^2(\xi,\zeta)\Big]
             + \frac{M^2}{{\cal P}^2}
             \frac{1}{\zeta}\,T_5^2(\xi,\zeta) 
             \Big]\bigg\}\,.
\nonumber\\
\end{eqnarray}
One may finally change to the basis in (\ref{eqD2}). The unpolarized diffractive structure functions
are then obtained by
\begin{eqnarray}
W_1^s(\beta,\eta)&=& U_1\\ 
W_2^s(\beta,\eta)&=&  U_2 - 2 \left(1 + \frac{1}{\eta}\right) U_3
+ \left(1+ \frac{1}{\eta}\right)^2 U_4 \\ 
W_3^s(\beta,\eta)&=&  U_2 + 2 
\left(1-\frac{1}{\eta}\right) U_3 +\left(1- \frac{1}{\eta}\right)^2 U_4 
\\ 
W_4^s(\beta,\eta)&=& -U_2 + \frac{2}{\eta} U_3                     
+ \left(1 -\frac{1}{\eta^2}\right) U_4~, 
\end{eqnarray}
with $U_i = U_i(\beta,\eta)$.
The structure functions $U_i$, resp. $W_i^s$, depend only on the measurable kinematic 
variables $\beta$ and $\eta$ and form the hadronic tensor (\ref{eqD2}) in 
the spin--averaged  case for general azimuthal angle and can be measured directly.
\section{The Antisymmetric Part of the Amplitude}
\label{sec-asy}
\renewcommand{\theequation}{\thesection.\arabic{equation}}
\setcounter{equation}{0}

\vspace{2mm}
\noindent
The Fourier transform for the antisymmetric
part of the amplitude again is taken over from Ref.~\cite{Geyer:2004bx}
(cf.~Eq. (5.2)):
\begin{align}
 T^\twz_{[\mu\nu]}\kln{q}  =
    \frac{1}{2} \sum_a \int D \mathbb{Z}\; \epsilon^{\mu\alpha\nu\beta}\,
    \frac{- q^2}{[(q\Pi)^2 - q^2 \Pi^2]}\,&
    \Bigg\{
    {q_{\alpha}\,{\cal K}^a_{5\beta}} F^{a}_{(5)\,1}(\mathbb{Z}) 
    + {q_{\alpha}\,\Pi_\beta} {(q{\cal K}_5^a)} F^{a}_{(5)\,2}(\mathbb{Z})
  \nonumber\\
& ~~    
    + {q_{\alpha}\,\Pi_\beta}{(\Pi{\cal K}_5^a)} F^{a}_{(5)\,3}(\mathbb{Z})
    \Bigg\}\frac{1}{R(1)+ \im \epsilon}\,,
    \label{Tas500}
\end{align}
with
\begin{eqnarray}
F^{(5)}_{a\,1}(\mathbb{Z})
&=&
    (q\Pi+\Pi^2)\,
    \int_0^1\frac{d\tau}{\tau^2}
            \;f_{5\,a}\Big(\frac{{\mathbb Z}}{\tau}\Big)
     + \Pi^2\,
     \int_0^1\frac{d\tau}{\tau^2} \int_0^1\frac{d\tau_1}{\tau^2_1}
            \;f_{5\,a}\Big(\frac{{\mathbb Z}}{\tau\tau_1}\Big)\,,
     \label{F51}
\end{eqnarray}\begin{eqnarray}
F^{(5)}_{a\,2}(\mathbb{Z})
&=&
    f_{5\,a}(\mathbb{Z})
    - \left(\frac{3\, q\Pi (q\Pi + \Pi^2)}{[(q\Pi)^2 - q^2 \Pi^2]}-2\right)
    \int_0^1\frac{d\tau}{\tau^2}
            \;f_{5\,a}\Big(\frac{{\mathbb Z}}{\tau}\Big)
\nonumber\\ & &   - \frac{3 \, (q\Pi)\,\Pi^2 }{[(q\Pi)^2 - q^2\Pi^2]}\,
   \int_0^1\frac{d\tau}{\tau^2} \int_0^1\frac{d\tau_1}{\tau^2_1}
            \;f_{5\,a}\Big(\frac{{\mathbb Z}}{\tau\tau_1}\Big)\,,
            \label{F52}\\
F^{(5)}_{a\,3}(\mathbb{Z})
&=&
    f_{5\,a}(\mathbb{Z})
    +  \frac{3 \, q\Pi (q^2 + q\Pi)}{[(q\Pi)^2 - q^2 \Pi^2]}\;
    \int_0^1\frac{d\tau}{\tau^2}
            \;f_{5\,a}\Big(\frac{{\mathbb Z}}{\tau}\Big)
\nonumber\\ & &    +\left(\!  \frac{3 \,  (q\Pi)^2}{[(q\Pi)^2 - q^2 \Pi^2]}-1\!\right)
    \int_0^1\frac{d\tau}{\tau^2} \int_0^1\frac{d\tau_1}{\tau^2_1}
            \;f_{5\,a}\Big(\frac{{\mathbb Z}}{\tau\tau_1}\Big)\,,
            \label{F53}
\end{eqnarray}
Let us remark that the symmetry $(\mathbb{Z}) \rightarrow -(\mathbb{Z})$ in 
$F_{a\,i}^{(5)}$ is a consequence of the symmetry in $f_{5a}$ as has been
shown in Ref.~\cite{Geyer:2004bx}, App. C. Consequently, 
all terms  $ \propto 1/{[R(0)+\im \epsilon]}$ drop out. 

We consider the imaginary part of this expression. Again,
only the terms  $\propto 1/[R(1)+ \im \epsilon]$ contribute, 
\begin{eqnarray}
 {\mathsf {Im}}\, T^\twz_{[\mu\nu]}\kln{q} 
 &= &   \frac{\pi}{4}\,\epsilon_{\mu\nu}^{\phantom{\mu\nu}\alpha\beta} \sum_a
    \int d\zeta \int_{1/\eta}^0 
\vartheta^2 \, 
    \frac{q^2}{(\vartheta\kappa)^3[(q{\cal P})^2 - q^2 {\cal P}^2]^{3/2}}\;
    \delta\big(\vartheta - \xi \big)
    \times
    \nonumber\\
    && \bigg\{
    {q_{\alpha}\,{\cal K}^a_{5\beta}}\, F^{(5)}_{a\,1}(\vartheta, \zeta)
    + \vartheta\kappa \,{q_{\alpha}\,{\cal P}_\beta} {(q{\cal K}_5^a)}\, F^{(5)}_{a\,2}(\vartheta, \zeta)
    + (\vartheta\kappa)^2{q_{\alpha}\,{\cal P}_\beta}{({\cal P}{\cal K}_5^a)}\, F^{(5)}_{a\,3}(\vartheta, 
\zeta)
   \bigg\}.
   \nonumber\\
   \label{Tas5}
\end{eqnarray}
It is convenient to introduce the distribution amplitudes $\Phi_{5a}{(i)}(\vartheta,\zeta)$
\begin{eqnarray}
    \Phi^{(0)}_{5\,a}(\vartheta,\zeta) &\equiv&  \vartheta\, f_{5\,a}(\vartheta,\zeta) \,,
\nonumber\\
\Phi^{(i)}_{5\,a}\left(\vartheta,\zeta\right)
    &=& \vartheta 
    \int^1_0\frac{d\tau_1}{\tau_1^2} \ldots
    \int^1_0\frac{d\tau_{i-1}}{\tau_{i-1}^2}
    \int^1_0\frac{d\tau_{i}}{\tau_{i}^2}\,
        f_{5\,a}\left(\frac{\vartheta}{\tau_1\cdots\tau_{i-1}\tau_i},\zeta\right)
    \nonumber
\\
    &=& \vartheta 
    \int^1_0\frac{d\tau_1}{\tau_1^2}\ldots
    \int^1_0\frac{d\tau_{i-1}}{\tau_{i-1}^2}
    \int^{{1/\eta }}_{\vartheta /(\tau_1\cdots\tau_{i-1})}\frac{dy_i}{y_i}\,
    \frac{\tau_1\cdots\tau_{i-1}\, y_i}{\vartheta }\,
         f_{5\,a}\left(y_i,\zeta\right)\,
    \nonumber\\
     &=& 
    \int^{{1/\eta }}_\vartheta \frac{dy_1}{y_1}\ldots
    \int^{{1/\eta }}_{y_{i-2}}\frac{dy_{i-1}}{y_{i-1}}
    \int^{{1/\eta }}_{y_{i-1}}\frac{dy_{i}}{y_{i}}\,
          \left[y_i\,f_{5\,a}\left(y_i,\zeta\right)\right]\,
    \nonumber\\
    &=& 
    \int^{{1/\eta }}_{\vartheta}\frac{dy}{y} \,
        \Phi^{(i-1)}_{5\,a}\left(y,\zeta\right)\,,
\label{NVF}
\end{eqnarray}
where a successive change of variables
$y_m={\vartheta}/({\tau_1\ldots\tau_{m-1}\tau_m}),\; m = n, \ldots, 1$,
has been made and the support restriction of
{$f_{5\,a}\left(\vartheta,\zeta\right)$} to ${{1/\eta}} \leq \vartheta \leq {{0}}$ 
has been observed.

Then, performing the $\vartheta$-integration with the help of the $\delta$-function,
one gets the following expression which simply may be obtained from 
\cite{Geyer:2004bx}, Eqs. (5.10)  replacing there $x \rightarrow -2\beta$:
\begin{align}
\label{Tas6}
{\mathsf{Im}}\, T^\twz_{[\mu\nu]}\kln{q}
    =& {\frac{\pi}{2}\,\epsilon_{\mu\nu}^{\phantom{\mu\nu}\alpha\beta}}\int d\zeta  \,
    \left(\frac{4\beta}{\xi}\,\frac{1}{[1+4\beta^2\mathcal{P}^2/Q^2]^{3/2}}\right)
    \times
\\
    &  \Bigg\{
    \frac{q_{\alpha}\,{\cal K}^a_{5\beta}}{q{\cal P}} 
    \left[\left(1 - \beta \xi\mathcal{P}^2/Q^2\right)
          \Phi_{5a}^{(1)}(\xi,\zeta)\,
    - \beta \xi\mathcal{P}^2/Q^2\,
          \Phi_{5a}^{(2)}(\xi,\zeta)\, \right]
\nonumber\\
    &  + \frac{q_{\alpha}\,\mathcal{P}_\beta}{q{\cal P}}
    \frac{(q{\cal K}_5^a)}{q{\cal P}}  
    \left[
          \Phi_{5a}^{(0)}(\xi,\zeta)
        - \frac{1 +2\beta \xi\,\mathcal{P}^2/Q^2}
                 {[1+4\beta^2\mathcal{P}^2/Q^2]^{1/2}}
          \Phi_{5a}^{(1)}(\xi,\zeta) \right.
\nonumber\\ & \left.        
+ 3\frac{\beta \xi \,\mathcal{P}^2/Q^2 }{[1+4\beta^2\mathcal{P}^2/Q^2]}
          \Phi_{5a}^{(2)}(\xi,\zeta)    \right]
\nonumber\\
\end{align}
\begin{align}
    &  - {\beta \xi}\frac{\mathcal{P}^2}{Q^2}
    \frac{q_{\alpha}\,\mathcal{P}_\beta}{q{\cal P}}
    \frac{(\mathcal{P}{\cal K}_5^a)}{\mathcal{P}^2}
    \left[   \Phi_{5a}^{(0)}(\xi,\zeta)
          -\frac{3}{[1+4\beta^2\mathcal{P}^2/Q^2]^{1/2}}
            \Phi_{5a}^{(1)}(\xi,\zeta) \right.
\nonumber\\ & \left.         + \frac{2- 4\beta^2 \mathcal{P}^2/Q^2}{[1+4\beta^2\mathcal{P}^2/Q^2]}
            \Phi_{5a}^{(2)}(\xi,\zeta)    \right]
    \Bigg\},
    \nonumber
\end{align}
This equation is written more conveniently as\footnote{The implicit 
$\eta$-dependence has been suppressed for brevity.}
\begin{eqnarray}
 {\mathrm{Im}}\, T^\twz_{[\mu\nu]}\kln{q}  &=& 
 -\pi \,\epsilon_{\mu\nu}^{\phantom{\mu\nu}\alpha\beta} \int d\zeta  \,
   \Bigg\{
    \frac{q_{\alpha}\,{\cal K}^a_{5\beta}}{q{\cal P}}
          \left[g_{a1}(\beta,\zeta) + g_{a2}(\beta,\zeta)\right]
    - \frac{q_{\alpha}\,\mathcal{P}_\beta}{q{\cal P}}
              \frac{(q{\cal K}_5^a)}{q{\cal P}} g_2(\beta,\zeta)
    \nonumber\\
   && \qquad\qquad\qquad\qquad
    + \frac{1}{2}\frac{\mathcal{P}^2}{Q^2}
              \frac{q_{\alpha}\,\mathcal{P}_\beta}{q{\cal P}}
              \frac{(\mathcal{P}{\cal K}_5^a)}{\mathcal{P}^2} g_{a0}(\beta,\zeta)
    \Bigg\}\,.
       \label{Tas7}
\end{eqnarray}
Here the following equalities
\begin{eqnarray}
\label{x1}
1-\beta\xi \mathcal{P}^2/Q^2 &=& \sqrt{1+4\beta^2\mathcal{P}^2/Q^2}= -(1+4\beta / \xi)\,,
\\
\label{x2}
\beta\frac{\partial}{\partial\beta}\,\xi &=&\frac{\xi}{\sqrt{1+4\beta^2\mathcal{P}^2/Q^2}}\,,
\\
\label{x3}
\beta\frac{\partial}{\partial\beta}\,\Phi_a^{(i)}(\xi;\zeta)&=&
-\frac{\xi}{\sqrt{1+4\beta^2\mathcal{P}^2/Q^2}}\,\Phi_a^{(i-1)}(\xi;\zeta)\,,
\\ 
    \beta \frac{\partial }{\partial\beta }\left(
    \frac{2\beta }{\xi}\,\frac{1}{[1+4\beta^2\mathcal{P}^2/Q^2]^{1/2}}\right)
    &=&
    \frac{2\beta }{\xi}\,
    \frac{1 - [1+ 4\beta^2\mathcal{P}^2/Q^2]^{1/2}}{[1+4\beta^2\mathcal{P}^2/Q^2]^{3/2}}
    =
    \frac{1}{2}\,
\frac{4\beta^2\mathcal{P}^2/Q^2 }{[1+4\beta^2\mathcal{P}^2/Q^2]^{3/2}}\,,
\nonumber\\
\end{eqnarray}
were used. For the functions $g_{ai}(\beta,\zeta)$ one finds the following expressions:
\begin{eqnarray}
   g_{a1}(\beta,\zeta)  
   &\equiv&
         -\beta\frac{\partial}{\partial \beta }
         \left[\beta\frac{\partial}{\partial \beta}\left(
        \frac{2\beta}{\xi}
        \frac{\Phi_{5a}^{(2)}(\xi,\zeta)}{[1+4\beta^2\mathcal{P}^2/Q^2]^{1/2}}
        \right)\right],
    \nonumber\\
    &=&
    \frac{-2\beta/\xi }{[1+4\beta^2\mathcal{P}^2/Q^2]^{3/2}}
    \bigg[\Phi_{5a}^{(0)}(\xi,\zeta)
    + 
    \frac{2\beta (2\beta -\xi)\mathcal{P}^2/Q^2}{[1+4\beta^2\mathcal{P}^2/Q^2]^{1/2}}
    \Phi_{5a}^{(1)}(\xi,\zeta)
\nonumber\\ &&    +
    \beta \xi \mathcal{P}^2/Q^2
    \frac{(2-4\beta^2\mathcal{P}^2/Q^2)}{[1+4\beta^2\mathcal{P}^2/Q^2]}
    \Phi_{5a}^{(2)}(\xi,\zeta)
    \bigg],
\end{eqnarray}
\begin{eqnarray}
    g_{a2}(\beta,\zeta) 
    &\equiv& 
    \beta
    \frac{\partial^2}{\partial \beta^2}\beta\left( \frac{2\beta}{\xi}
    \frac{\Phi_{5a}^{(2)}(\xi,\zeta)}
    {[1+4\beta^2\mathcal{P}^2/Q^2]^{1/2}}
    \right)
    \nonumber\\
    &=&
    \frac{2\beta/\xi }{[1+4 \beta^2\mathcal{P}^2/Q^2]^{3/2}}
    \left[
    \Phi_{5a}^{(0)}(\xi,\zeta)
    -
    \frac{1 +2\beta \xi \mathcal{P}^2/Q^2}{[1+4\beta^2\mathcal{P}^2/Q^2]^{1/2}}
    \Phi_{5a}^{(1)}(\xi,\zeta)
\right.
\nonumber\\ &&  \left.  +
    3 \frac{\beta \xi \mathcal{P}^2/Q^2}{[1+\beta^2\mathcal{P}^2/Q^2]}
    \Phi_{5a}^{(2)}(\xi,\zeta)
    \right],   
\\
    g_{a1}(\beta,\zeta) + g_{a2}(\beta,\zeta) &\equiv&
     \beta\frac{\partial}{\partial \beta}\left(
    \frac{2\beta}{\xi}
    \frac{\Phi_{5a}^{(2)}(\xi,\zeta)}
    {[1+4\beta^2\mathcal{P}^2/Q^2]^{1/2}}
    \right)
    \nonumber\\
    &=& 
    \frac{{-2\beta}/{\xi}}{[1+x^2\mathcal{P}^2/Q^2]^{3/2}}
    \left[\left(1 -\beta\xi\mathcal{P}^2/Q^2\right)
    \Phi_{5a}^{(1)}(\xi,\zeta)\,
    -\beta\xi\mathcal{P}^2/Q^2\,
    \Phi_{5a}^{(2)}(\xi,\zeta)\,
    \right],
  \nonumber\\ 
\end{eqnarray}\begin{eqnarray}
    g_{a0}(\beta,\zeta) &=& \beta^2 \frac{\partial^2}{\partial \beta^2}
    \left( 4\beta^2
    \frac{\Phi_{5a}^{(2)}(\xi,\zeta)}
    {[1+4\beta^2\mathcal{P}^2/Q^2]^{1/2}}
    \right)
    \nonumber\\
    &=& \frac{4\beta^2}{[1+4\beta^2 \mathcal{P}^2/Q^2]^{3/2}}
    \left[
    \Phi_{5a}^{(0)}(\xi,\zeta)
    -
    \frac{3}{[1+4\beta^2\mathcal{P}^2/Q^2]^{1/2}}
    \Phi_{5a}^{(1)}(\xi,\zeta)
\right.
\nonumber\\ &&    +\left.
    \frac{2- 4\beta^2 \mathcal{P}^2/Q^2}{[1+4\beta^2\mathcal{P}^2/Q^2]}
    \Phi_{5a}^{(2)}(\xi,\zeta)
    \right].
    \label{G_rel}
\end{eqnarray}
%
Obviously, the contributions to the various kinematic
expressions are related to only one function. Namely,
introducing 
\begin{eqnarray}
\label{calF}
\mathcal{F}_a(\beta,\eta,\zeta)
    =  \frac{-2\beta}{\xi} \frac{\Phi_{5a}^{(2)}(\xi,\zeta)}
       {[1+4\beta^2\mathcal{P}^2/Q^2]^{1/2}}  
\end{eqnarray}
we may rewrite the functions $g_{ai}$ as follows
\begin{eqnarray}
    g_{a1}(\beta,\eta,\zeta) 
    &=&    
    \beta\frac{\partial}{\partial \beta}
    \beta\frac{\partial}{\partial \beta} 
    \mathcal{F}_a(\beta,\eta,\zeta)
    \nonumber\\
    g_{a2}(\beta,\eta,\zeta) 
    &=& 
    - \beta\frac{\partial}{\partial \beta}
    \Big(\beta\frac{\partial}{\partial \beta} + 1\Big) 
    \mathcal{F}_a(\beta,\eta,\zeta)
    \nonumber\\
    g_{a 0}(\beta,\eta,\zeta) 
    &=&  
    -2\, \beta\frac{\partial}{\partial \beta}
    \Big(\beta\frac{\partial}{\partial \beta} - 1\Big) 
    \beta\,\xi\,\mathcal{F}_a(\beta,\eta,\zeta)\,.
\label{G_rel2}
\end{eqnarray}
One finally has to perform the $\zeta$--integral to obtain the
diffractive structure functions in the polarized case. The distribution
functions $\left. g_{ai}(\beta,\eta,\zeta)\right|_{i=0}^2$ occur in products
with $\left. \zeta^k\right|_{k=0}^2$ and result into six structure functions  
of the form for each invariant $a$
\begin{eqnarray}
\label{eqGG}
    G_{aj}(\beta,\eta) &=& \int d\zeta g_{aj}(\beta,\eta,\zeta), \\
    H_{aj}(\beta,\eta) &=& \int d\zeta\, \zeta  g_{aj}(\beta,\eta,\zeta),\\
    K_{aj}(\beta,\eta) &=& \int d\zeta\, \zeta^2 g_{aj}(\beta,\eta,\zeta),
\end{eqnarray}
leading to
\begin{eqnarray}
 {\mathrm{Im}}\, T^\twz_{[\mu\nu]}\kln{q}  &=& -\pi  \,
             \epsilon_{\mu\nu}^{\phantom{\mu\nu}\alpha\beta}\,
   \Bigg\{ 
    \frac{q_{\alpha}\,{\cal K}^a_{5\beta}}{qp_-}
          \big(G_{a1}(\beta,\eta) + G_{a2}(\beta,\eta)\big)
    \nonumber\\
   && \qquad\qquad
    - \frac{q_{\alpha}\,p_{-\beta}}{qp_-}\left(
              \frac{q{\cal K}_5^a}{qp_-} G_{a2}(\beta,\eta)
   - \frac{1}{2}\frac{p_-{\cal K}_5^a}{Q^2} G_{a0}(\beta,\eta)
    - \frac{1}{2}
        \frac{\pi_- {\cal K}_5^a}{Q^2} H_{a0}(\beta,\eta)
   \right)
    \nonumber\\
    && \qquad \qquad 
    -
    \frac{q_{\alpha}\,\pi_{-\beta}}{qp_-}\left(
              \frac{q{\cal K}_5^a}{qp_-} H_{a2}(\beta,\eta)
    - \frac{1}{2}\frac{p_-{\cal K}_5^a}{Q^2} H_{a0}(\beta,\eta)
    - \frac{1}{2}\frac{\pi_-{\cal K}_5^a}{Q^2} K_{a0}(\beta,\eta)\right)   
     \Bigg\}\,. \nonumber
 \label{Tas8}
\end{eqnarray}
Inserting the three invariants ${\cal K}_{a5}$ explicitly one obtains
\begin{align}
\label{ergas}
{\mathrm{Im}}\, T^\twz_{[\mu\nu]}\kln{q}  
    &= 
    -\pi \, \epsilon_{\mu\nu}^{\phantom{\mu\nu}\alpha\beta}\, 
    \Bigg\{
    \frac{q_{\alpha}\,S^{\mathrm T}_{\beta}}{qp_-}
          \big(G_{1\,1} + G_{1\,2}\big)  
    \nonumber \\
&   + \frac{q_{\alpha}\,p^{\mathrm T}_{-\beta}}{qp_-}
    \Bigg[
    -\frac{q S}{q p_-}  G_{1\,2}
+ \frac{p_2 S}{M^2} 
    \bigg[
    G_{2\,1}
   \nonumber \\
&   \qquad
    +\frac{1}{2Q^2} \left({p_-^2 }  G_{2\,0}
    + {p_-\pi_-} ( G_{3\,0}+H_{2\,0} ) 
  +{\pi_-^2} H_{3\,0} \Big) 
  +M^2 G_{1\,0} + M^2\frac{\eta -1}{\eta}  H_{1\,0} \right)\bigg]
     \Bigg] 
     \nonumber \\
&  + \frac{q_{\alpha}\,\pi^{\mathrm T}_{-\beta}}{qp_-}
    \Bigg[
    - \frac{q S}{q p_-} H_{1\,2}(\beta,\eta)             
  + \frac{p_2 S  }{M^2}
  \bigg[G_{3\,1} + G_{3\,2} - H_{2\,2}
  \nonumber \\
&   \qquad
    +\frac{1}{2Q^2}  
    \left({p_-^2 }  H_{2\,0} 
    +{p_-\pi_-} (H_{3\,0} + K_{2\,0}) 
    + {\pi_-^2} K_{3\,0} 
    + M^2 H_{1\,0} + M^2\frac{\eta -1}{\eta} K_{1\,0} \right)
    \bigg]  
    \Bigg]
    \Bigg\}, 
\end{align}
with 
\begin{eqnarray}
p_-^2 = t, \qquad
p_-\pi_- = - t/\eta, \qquad
\pi_-^2 = 4M^2 - t(1-1/\eta^2), \qquad \frac{\eta - 1}{\eta} = \frac{2 \beta}{x}~.
\end{eqnarray}
This structure of the hadronic tensor specifies the general structure 
obtained for three characteristic 4--vectors $p_1, p_2$ and $S$ in 
Eq.~(\ref{eqH2}) for the present process. One may cast (\ref{ergas}) into the 
form of (\ref{eqH2}) using Shouten identities \cite{SHOUT}:
\begin{eqnarray}
X_{\mu} \varepsilon_{\nu \rho \sigma \tau}
&=&
 X_{\nu} \varepsilon_{\mu \rho \sigma \tau}
+X_{\rho} \varepsilon_{\nu \mu \sigma \tau}
+X_{\sigma} \varepsilon_{\nu \rho \mu  \tau}
+X_{\tau} \varepsilon_{\nu \rho \sigma \mu}
\\
g_{\lambda\mu} \varepsilon_{\nu \rho \sigma \tau}
&=&
  g_{\lambda\nu} \varepsilon_{\mu \rho \sigma \tau}
+ g_{\lambda\rho} \varepsilon_{\nu \mu \sigma \tau}
+ g_{\lambda\sigma} \varepsilon_{\nu \rho \mu \tau}
+ g_{\lambda\tau} \varepsilon_{\nu \rho \sigma \mu}~.
\end{eqnarray}

Note that the structure functions $J_{k0},~~J = G, H, K$ are suppressed by 
$\mu^2/Q^2$ relative to the other contributions, with $\mu^2$ a hadronic 
scale. In the Bjorken limit, these structure functions do not contribute.
\section{Relations between Diffractive Structure Functions}
\renewcommand{\theequation}{\thesection.\arabic{equation}}
\setcounter{equation}{0}
Relations between different structure functions as observed in deep--inelastic 
scattering
\cite{Callan:1969uq,Wandzura:1977qf,Blumlein:1996vs,Blumlein:1998nv,Blumlein:2000cx} hold also for 
polarized and 
unpolarized 
diffractive scattering
in the limit $M^2, t \rightarrow 0$, cf. \cite{Blumlein:2001xf,Blumlein:2002fw}. 
The situation is 
more involved for the case studied in the present paper, since the structure functions
emerge as a $\zeta-$integral of sub-system structure functions, which accounts for the
two--particle nature of the wave--function. Thus the corresponding relations can be established
for the un-integrated $\zeta$-dependent functions only.  
\subsection{Unpolarized Case}

\vspace{2mm}
\noindent
As before in the case of deep-inelastic scattering 
\cite{Georgi:1976ve,Blumlein:1998nv} and non-forward scattering 
\cite{Geyer:2004bx} we seek for a representation of the 
distribution functions ${W}^{\rm diff}_{1,2,L}, V^{\rm {diff}}_{a\,(0,1)}$ in terms of 
a single function. One may choose the distribution functions
$\Phi_a^{(2)}(\xi;\zeta)$ (\ref{eqphia2}) and express 
the functions ${W}^{\rm diff}_{1,2,L}, V^{\rm {diff}}_{a\,(0,1)}$
through the following differential relations~:
\begin{eqnarray}
{W}^{\rm diff}_{a\, \rm L}(\xi(\beta),\beta,\eta;\zeta)
 &=&  \frac{ {\cal P}^2}{Q^2}\;\beta^2 \frac{\pd}{\pd \beta}
 \bigg(\frac{-2\beta}{\xi(\beta)}
 \frac{ \Phi_a^{(2)}(\xi(\beta);\zeta)}{\sqrt{1+4\beta^2 {\cal P}^2/Q^2}}\bigg),
 \label{WL}\\
 \frac{2q{\cal P}}{{\cal P}^2}\, {W}^{\rm diff }_{a\,2}(\xi(\beta),\beta,\eta;\zeta)
 &=& \beta^2
 \frac{\pd^2}{\pd \beta^2}
 \bigg(\frac{4\beta^2}{\xi(\beta)^2}
 \frac{\Phi_a^{(2)}(\xi(\beta);\zeta)}{\sqrt{1+4\beta^2 {\cal P}^2/Q^2}}\bigg),
\label{W2}
\end{eqnarray}
Note that the longitudinal distribution function ${W}^{\rm diff}_{a\, \rm 
L}(\xi(\beta),\beta,\eta;\zeta)$ vanishes in the limit ${\cal P}^2/Q^2~\rightarrow~0$.

A generalized Callan--Gross relation, which holds for diffractive scattering in the
limit $M^2,~t~\rightarrow~0$, \cite{Blumlein:2001xf}, 
is broken as in the case of deep--inelastic 
scattering \cite{Georgi:1976ve}. The distribution functions ${W}_{(1,2)}^{\rm diff}$ 
are related by
\begin{eqnarray}
 {W}^{\rm diff }_{a\,1}(\xi(\beta),\beta,\eta;\zeta)
+{W}^{\rm diff }_{a\,\rm L}(\xi(\beta),\beta,\eta;\zeta)
 &=&
 \frac{(1+4\beta^2 {\cal P}^2/Q^2)}{(-4\beta)}\,
 \frac{2q{\cal P}}{{\cal P}^2}\,{W}^{\rm diff }_{a\,2}(\xi(\beta),\beta,\eta;\zeta) 
\nonumber\\ &=&
\frac{({\cal P}^{\mathrm T})^2}{{\cal P}^2}\,
 {W}^{\rm diff }_{a\,2}(\xi(\beta),\beta,\eta;\zeta)\,~.
 \label{W1}
\end{eqnarray}
The distribution functions $V^{\rm {diff}}_{a\,(0,1)}$ can be expressed by the distribution functions
${W}_{1,L}^{\rm diffr}$ directly 
\begin{eqnarray}
  2 V^{\rm {diff}}_{a\,0}(\xi(\beta),\beta,\eta;\zeta) 
&=& 
    W^{\rm {diff}}_{a\,L}(\xi(\beta),\beta,\eta;\zeta)
- 2 W^{\rm {diff}}_{a\,1}(\xi(\beta),\beta,\eta;\zeta)\,,
\label{V0x}\\
 \frac{1}{2}\,V^{\rm {diff}}_{a\,1}(\xi(\beta),\beta,\eta;\zeta)
 &=&
\sqrt{1+4\beta^2{\cal P}^2/Q^2}\,
 \beta\frac{\pd}{\pd \beta}\,
 W^{\rm {diff}}_{a\,L}(\xi(\beta),\beta,\eta;\zeta)
\nonumber\\ & &
 + \bigg(1 - \frac{2}{\sqrt{1+4\beta^2{\cal P}^2/Q^2}}\bigg)
 \,W^{\rm {diff}}_{a\,L}(\xi(\beta),\beta,\eta;\zeta)
  \nonumber\\
 &&
   - \;
 \frac{4\beta^2 {\cal P}^2/Q^2}{[1+4\beta^2{\cal P}^2/Q^2]^{3/2}}
 \int_{\beta}^1 \frac{d \rho}{\rho^2}\,
 W^{\rm {diff}}_{a\,L}(\xi(\beta\rho),\beta\rho,\eta;\zeta)\, .
\label{V1x}
\end{eqnarray}
Three of the above distribution functions are independent.
To obtain the four diffractive structure functions on the level of 
observables $W_i^s|_{i=1}^4$
in the unpolarized case (\ref{eqD2}) the $\zeta$--integral has to be performed.
As shown in section~\ref{sec-unp} the respective linear combinations are
weighted by different $\zeta$--dependent functions, that in general no 
relations exist on the level of structure functions.
\subsection{Polarized Case}

\vspace{2mm}
\noindent
Below the $\zeta$--integral the distribution functions $g_{ai}|_{i=0}^2$
are all functions of ${\cal F}_a(\beta,\eta,\zeta)$ (\ref{calF}). Therefore one 
may express the functions $g_{a0}$ and $g_{a2}$ in terms of $g_{a1}$ as the central 
function. This is achieved applying the relations 
\begin{eqnarray}
    \beta\frac{\partial}{\partial \beta}\mathcal{F}_a(\beta,\eta,\zeta)  &=&
    - \int_\beta^1 \frac{dy}{y}g_{a1}^{\rm tw2}(y,\eta,\zeta)\,, \\
    \mathcal{F}_a(\beta,\eta,\zeta)    &=&
    - \int_\beta^1 \frac{dy}{y}
      \int_y^{1} \frac{dy'}{y'}g_{a1}^{\rm tw2}(y',\eta,\zeta)\,.
\end{eqnarray}
We consider twist--2 operators only and find
the Wandzura-Wilczek relation 
\begin{eqnarray}
\label{ww11}
    g^{\twz}_{a2}(\beta,\eta,\zeta)  
    &=& 
    -\, g_{a1}^{\rm tw2}(\beta,\eta,\zeta) +
    \int_\beta^1 \frac{dy}{y}g_{a1}^{\rm tw2}(y,\eta,\zeta)\,
    \label{g22}
\end{eqnarray} 
between $g_{a2}^{\rm tw2}$ and $g_{a1}^{\rm tw2}$. All target mass and 
$t$--corrections are
absorbed into the structure functions. Note that this relation holds for 
all invariants $a$ independently. The validity of the Wandzura--Wilczek 
relation also in case of diffractive scattering at general hadronic scales
$M^2, t$ is a further example in a long list of cases: 
covariant parton model 
\cite{Roberts:1996ub,Blumlein:1996vs,Blumlein:1996tp},
target- and quark--mass corrections \cite{Piccione:1997zh,Blumlein:1998nv}
gluon-induced heavy flavor production \cite{Blumlein:2003wk},
non--forward scattering \cite{Blumlein:2000cx},
diffractive scattering in the limit $M^2, t \rightarrow 0$ 
\cite{Blumlein:2002fw}.

The distribution function $g_{a0}^{\rm tw2}$ obeys the relation
\begin{eqnarray}
    g_{a0}^{\twz}(\beta,\eta,\zeta)  
    &=& 
    -\,2\,\beta\xi\,g_{a1}^{\rm tw2}(\beta,\eta,\zeta)
    - \frac{8\beta^2-2\beta\xi}{[1+4\beta^2\mathcal{P}^2/Q^2]^{1/2}}
    \int_\beta^1\!
    \frac{dy}{y}g_{a1}^{\rm tw2}(y,\eta,\zeta)
    \nonumber \\
    & &    
    -\frac{8\beta^2}{[1+4\beta^2\mathcal{P}^2/Q^2]^{3/2}}
    \int_\beta^1\! \frac{dy}{y} \!
    \int_y^1\! \frac{dy'}{y'}
    g_{a1}^{\rm tw2}(y',\eta,\zeta)\,.
    \label{g02}
\end{eqnarray}
It contributes to parts of the hadronic tensor which are suppressed by 
an overall factor $\mu^2/Q^2$, with $\mu$ a hadronic mass scale.
In the related case of non--forward scattering the emergence of this 
function has been observed in Ref.~\cite{Blumlein:2000cx}.

For $g_{a2}$ the $\zeta$--integral can be performed maintaining
the structure of relation (\ref{ww11}). Therefore the Wandzura--Wilczek 
relation also holds for the diffractive structure functions 
\begin{eqnarray}
    G^{\twz}_{a2}(\beta,\eta)  &=& -\, G_{a1}^{\rm tw2}(\beta,\eta) +
                \int_\beta^1 \frac{dy}{y}G_{a1}^{\rm tw2}(y,\eta)\,,
    \label{G22}
    \\
    H^{\twz}_{a2}(\beta,\eta)  &=& -\, H_{a1}^{\rm tw2}(\beta,\eta) +
                \int_\beta^1 \frac{dy}{y}H_{a1}^{\rm tw2}(y,\eta)\,.
    \label{H22}
\end{eqnarray}
Due to the $\zeta$--dependence of $g_{a0}$ through $\xi$ and ${\cal P}^2$
one needs to know $g_{a1}(\beta,\eta,\zeta)$ to calculate the structure 
functions $J_{a,0}(\beta,\eta)$. As the integral (\ref{eqGG}) cannot be 
inverted, relations between the structure functions $J_{a,0}$ and the 
other polarized structure functions cannot be established unless referring
to the un-integrated distribution functions $g_{a1}(\beta,\eta,\zeta)$.
\section{Conclusions}
\renewcommand{\theequation}{\thesection.\arabic{equation}}
\setcounter{equation}{0}

\vspace{2mm}
\noindent
Deep--inelastic diffractive scattering, like other hard scattering 
processes off nucleons, requires target mass corrections in the region of 
lower scales of $Q^2$. In fact the nucleon mass $M$ is not the only 
hadronic scale relevant to this process in which both the incoming and 
outgoing nucleon play a role. The modulus of the invariant $t = (p_2 - 
p_1)^2$ on average is of the same size as $M^2$.~\footnote{In case of 
related semi--exclusive processes in which more than one final--state 
hadron is well separated in rapidity from the inclusively produced hadrons
other invariants more would emerge.} It appears to be natural to seek for
a formalism dealing with both the target mass and $t$--effects 
simultaneously. Diffractive factorization and the use of A. Mueller's 
generalized optical theorem made it possible to reformulate diffractive 
scattering in terms of deep--inelastic scattering off an effective 
two--nucleon state accounting for $t$. We applied the formalism 
of the (non-local) light-cone expansion in the case of non--forward 
scattering and adapted it to the kinematics present in deep--inelastic 
diffractive scattering. The formalism accounting for target 
mass and $t$--corrections was then implemented generalizing the 
picture derived in the 
limit $M^2, t \rightarrow 0$ in earlier work both for the unpolarized  
and polarized case. The present formalism can in principle be extended to 
the case of higher twist concerning the corresponding operator expressions 
$\Omega_\alpha^{(5)}$. However, besides factorizable contributions also 
non-factorizable terms have to be described.

While in the limit $M^2, t \rightarrow 0$ all diffractive structure 
functions can be expressed in terms of diffractive parton densities,
this is no longer the case at lower scales of $Q^2$ in regions where 
$M^2$ and $t$--corrections become relevant. Here we mean the {\sf kinematic}
$t$ contributions which have to be distinguished from the dynamic  
contributions emerging in the non--perturbative distribution functions.
The reason for this is that the 2--particle kinematics of the incoming and 
outgoing nucleon deviates from  quasi--collinearity which was instrumental 
to unify the boost variables of both particles to a single one. The 
condition for the absorptive part is now related to a 
$\delta$--distribution with a more complicated argument leaving the 
integral over the relative transverse motion of the two nucleons.
This integral cannot be inverted in general referring to the accessible 
kinematic variables. The corresponding distribution functions
$\Phi_{a(5)}(\xi,\beta,\eta,\zeta;t)$ have to be determined by 
non--perturbative methods in the future.

Although the partonic picture does not hold for the diffractive structure 
functions one still finds it below the $\zeta$--integral. At the 
level of twist--2 the structure functions can be built from the
corresponding operator expectation values as in the case of 
deep--inelastic scattering since the specifics of diffractive scattering is 
moved into the corresponding two--particle wave--functions. Consequently,
the logarithmic scaling violations, which can be completely associated 
with that of the {\sf operators}, 
cf.~\cite{Blumlein:1999sc,Blumlein:2001xf}, are found to be the same as in 
diffractive scattering changing the variables $x$ in the deep--inelastic 
case to $\beta$ in the diffractive case.

The presence of target mass and $t$--effects enlarges the number of 
structure functions determining the hadronic tensor. In the unpolarized case
four structure functions contribute, which cannot be related to each other
directly. The Callan--Gross relation does not hold.  
However, in the polarized case the Wandzura--Wilzcek relation remains unbroken
and holds even separately for the contributions to the different invariants
${\cal K}_a|_{a=3}^5$. In addition to the structure functions surviving
in the limit $M^2, t \rightarrow 0, Q^2 \rightarrow \infty$ several
new structure functions contribute. They are damped however $\propto 1/Q^2$
w.r.t. the other contributions.
The present formalism can be used in experimental analysis of deep--inelastic 
diffractive scattering data referring to suitable models for the un--integrated
distribution functions depending on $\zeta$, for which rigorous determination
using methods of non--perturbative QCD do not yet exits. In this way the structures
being derived in the present paper can be tested.

\vspace{2mm}\noindent
{\bf Acknowledgments:} This work was supported in part by DFG
Sonderforschungsbereich Transregio 9, Computergest\"utzte Theoretische
Physik.


\end{document}